\documentclass[twocolumn,pre,aps,showpacs,showkeys,amsmath]{revtex4}
\usepackage{graphicx}

\begin{document}

\title{Effect of Spike-Timing-Dependent Plasticity on Stochastic Burst Synchronization in A Scale-Free Neuronal Network}

\author{Sang-Yoon Kim}
\email{sykim@icn.re.kr}
\author{Woochang Lim}
\email{wclim@icn.re.kr}
\affiliation{Institute for Computational Neuroscience and Department of Science Education, Daegu National University of Education, Daegu 42411, Korea}

\begin{abstract}
We consider an excitatory population of subthreshold Izhikevich neurons which cannot fire spontaneously without noise. As the coupling strength passes a threshold, individual neurons exhibit noise-induced burstings.
This neuronal population has adaptive dynamic synaptic strengths governed by the spike-timing-dependent plasticity (STDP). However, STDP was not considered in previous works on stochastic burst synchronization (SBS)
between noise-induced burstings of sub-threshold neurons. Here, we study the effect of additive STDP on SBS by varying the noise intensity $D$ in the Barab\'{a}si-Albert scale-free network (SFN). One of our main findings
is a “Matthew” effect in synaptic plasticity which occurs due to a positive feedback process. Good burst synchronization (with higher bursting measure) gets better via long-term potentiation (LTP) of synaptic strengths,
while bad burst synchronization (with lower bursting measure) gets worse via long-term depression (LTD). Consequently, a step-like rapid transition to SBS occurs by changing $D$, in contrast to a relatively smooth
transition in the absence of STDP. We also investigate the effects of network architecture on SBS by varying the symmetric attachment degree $l^*$ and the asymmetry parameter $\Delta l$ in the SFN, and Matthew effects are
also found to occur by varying $l^*$ and $\Delta l$. Furthermore, emergences of LTP and LTD of synaptic strengths are investigated in details via our own microscopic methods based on both the distributions of time delays
between the burst onset times of the pre- and the post-synaptic neurons and the pair-correlations between the pre- and the post-synaptic IIBRs (instantaneous individual burst rates). Finally, a multiplicative STDP case
(depending on states) with soft bounds is also investigated in comparison with the additive STDP case (independent of states) with hard bounds. Due to the soft bounds, a Matthew effect with some quantitative differences
is also found to occur for the case of multiplicative STDP.
\end{abstract}


\maketitle

\section{Introduction}
\label{sec:INT}Recently, brain rhythms in health and disease have attracted much attention \cite{Buz,TW,Rhythm1,Rhythm2,Rhythm3,Rhythm4,Rhythm5,Rhythm6,Rhythm7,Rhythm8,Rhythm9,Rhythm10,Rhythm11,Rhythm12,Rhythm13}. These brain rhythms appear through synchronization between individual firings in neural circuits.
Population synchronization of neural firing activities may be used for efficient sensory and cognitive processing (e.g., feature integration, selective attention, and memory formation) \cite{W_Review,Gray}, and
it is also correlated with pathological rhythms associated with neural diseases (e.g., epileptic seizures and tremors in the Parkinson's disease) \cite{ND1,ND2}. This kind of neural synchronization has been
intensively studied for the case of suprathreshold neurons exhibiting spontaneous regular firings like clock oscillators \cite{W_Review}. On the other hand, the case of subthreshold neurons (which cannot fire
spontaneously) has received little attention. With the help of noise, subthreshold neurons exhibit irregular firings like Geiger counters. Here, we are concerned about neural synchronization between
noise-induced firings.

Noise-induced firing patterns of subthreshold neurons were investigated in many physiological and pathophysiological aspects \cite{Braun1}. For example, sensory receptor neurons were found to use noise-induced firings
for encoding environmental electric or thermal stimuli, which are generated through the ``constructive'' interplay of subthreshold oscillations and noise \cite{Braun2,Longtin1}. A distinct characteristic of noise-induced
firings is occurrence of ``skipping'' of spikes at random integer multiples of a basic oscillation period (i.e., occurrence of stochastic phase locking) \cite{Braun1,Braun2,Longtin1,Longtin2}. These noise-induced firings
of a single subthreshold neuron become most coherent at an optimal noise intensity, which is called coherence resonance (or autonomous stochastic resonance without periodic forcing) \cite{Longtin2}. Furthermore,
array-enhanced coherence resonance was also found to occur in an ensemble of subthreshold neurons \cite{CR1,CR2,CR3,CR4,CR5}. In this way, noise may play a constructive role in the emergence of dynamical order in
certain circumstances.

Particularly, we are interested in noise-induced firings of subthreshold bursting neurons. There are several representative bursting neurons; for example, intrinsically bursting neurons and chattering neurons in the
cortex \cite{CT1,CT2}, thalamic relay neurons and thalamic reticular neurons in the thalamus \cite{TRN1,TRN2,TR}, hippocampal pyramidal neurons \cite{HP}, Purkinje cells in the cerebellum \cite{PC}, pancreatic
$\beta$-cells \cite{PBC1,PBC2,PBC3}, and respiratory neurons in the pre-B\"{o}tzinger complex \cite{BC1,BC2}. Due to a repeated sequence of spikes in the bursting, there are many hypotheses on the importance of bursting
activities in neural computation \cite{Burst2,Izhi2,Burst4,Burst5,Burst6}; for example, (a) bursts are necessary to overcome the synaptic transmission failure, (b) bursts are more reliable than single spikes in evoking
responses in post-synaptic neurons, (c) bursts evoke long-term potentiation/depression (and hence affect synaptic plasticity much greater than single spikes), and (d) bursts can be used for selective communication between neurons. As is well known, burstings occur when neuronal activity alternates, on a slow timescale, between a silent phase and an active (bursting) phase of fast repetitive spikings
\cite{Burst2,Izhi,Burst1,Rinzel1,Rinzel2,Burst3}. This kind of bursting activity occurs due to the interplay of the fast ionic currents leading to spiking activity and the slower currents modulating the spiking activity. Thus, the dynamics of bursting neurons have two timescales: slow bursting timescale and fast spiking timescale. Consequently, bursting neurons exhibit two different patterns of synchronization due to the slow and the fast timescales of bursting activity: burst synchronization (synchrony on the slow bursting timescale) which characterizes a temporal coherence between the (active phase) burst onset times (i.e., times at which burstings start in active phases) and spike synchronization (synchrony on the fast spiking timescale) which refers to a temporal coherence between intraburst spikes fired by bursting neurons in their respective active phases \cite{Burstsync1,Burstsync2}. Recently, burst and spike synchronizations have been studied in many aspects  \cite{BSsync1,BSsync2,BSsync3,BSsync4,BSsync5,BSsync6,BSsync7,BSsync8,BSsync9,BSsync10,BSsync11,BSsync12,BSsync13,BSsync14,BSsync15,BSsync16,BSsync17,BSsync18,BSsync19,BSsync20,BSsync21,BSsync22,BSsync23,BSsync24,BSsync25,
BSsync26}. However, most of these studies were focused on the suprathreshold case, in contrast to subthreshold case of our concern.

Here, we study stochastic burst synchronization (SBS) (i.e. population synchronization between noise-induced burstings of subthreshold neurons) which may be associated with brain functions of encoding
sensory stimuli in the noisy environment. Recently, such SBS has been found to occur in an intermediate range of noise intensity through competition between the constructive and the destructive roles of noise \cite{Kim1,Kim2}.
As the noise intensity passes a lower threshold, a transition to SBS occurs due to a constructive role of noise stimulating coherence between noise-induced burstings. However, when passing a higher threshold,
another transition from SBS to desynchronization takes place due to a destructive role of noise spoiling the SBS.
We note that synaptic coupling strengths were static in the previous works on SBS \cite{Kim1,Kim2}. However, in real brains synaptic strengths may be potentiated \cite{LTP1,LTP2,LTP3} or depressed \cite{LTD1,LTD2,LTD3,LTD4}
for adaptation to the environment. These adjustments of synapses are called the synaptic plasticity which provides the basis for learning, memory, and development \cite{Abbott1}.
In contrast to previous works where synaptic plasticity was not considered \cite{Kim1,Kim2}, as to the synaptic plasticity, we consider a Hebbian spike-timing-dependent plasticity (STDP) \cite{EtoE1,EtoE2,EtoE3,EtoE4,EtoE5,EtoE7,EtoE8,BTDP1,BTDP2,STDP1,STDP2,STDP3,STDP4,STDP5,STDP6,STDP7,STDP8}. For the STDP, the synaptic strengths change through a Hebbian plasticity rule depending on the relative time
difference between the pre- and the post-synaptic burst onset times. When a pre-synaptic burst precedes a post-synaptic burst, long-term potentiation (LTP) occurs; otherwise, long-term depression (LTD) appears.
Through the process of LTP and LTD in synaptic strengths, STDP controls the efficacy of diverse brain functions.
Many models for STDP have been employed to explain results on synaptic modifications occurring in diverse neuroscience topics for health and disease (e.g., temporal sequence learning \cite{TSLearning}, temporal pattern recognition \cite{EtoE6}, coincidence detection \cite{EtoE0}, navigation \cite{Navi}, direction selectivity \cite{DirSel}, memory consolidation \cite{Memory}, competitive/selective development \cite{Devel}, and deep brain stimulation \cite{iSTDP2}). Recently, the effects of STDP on population synchronization for the case of coupled (spontaneously-firing) suprathreshold neurons were studied in various aspects \cite{Brazil1,Brazil2,Tass1,Tass2}, and in the case of subthreshold spiking neurons (which cannot fire spontaneously without noise) stochastic spike synchronization (i.e., population synchronization between noise-induced spikings) was also studied in a small-world network with STDP \cite{SSS}.

In this paper, we consider an excitatory population of subthreshold Izhikevich neurons \cite{Izhi1,Izhi2,Kim1}. As the coupling strength passes a threshold, individual neurons exhibit noise-induced burstings. In the absence of STDP, SBS between noise-induced burstings of subthreshold neurons for the globally-coupled case was found to occur over a large range of intermediate noise intensities through competition between the constructive and the destructive roles of noise, as shown in our previous work \cite{Kim1}. Here, we investigate the effect of additive STDP (independent of states) on the SBS by varying the noise intensity $D$ in the Barab\'{a}si-Albert scale-free network (SFN) with symmetric preferential attachment with the same in- and out-degrees [$l_{in} = l_{out} = l^*~(=10)]$ \cite{BA1,BA2}, and compare its results with those in the absence of STDP. This type of SFNs exhibit a power-law degree distribution (i.e., scale-free property), and hence they become inhomogeneous ones with a few ``hubs'' (i.e., super-connected nodes), in contrast to statistically homogeneous networks such as random graphs and small-world networks. One of our main findings is a Matthew effect in synaptic plasticity which occurs due to a positive feedback process, similar to the case of stochastic spike synchronization \cite{SSS}.
Good burst synchronization with higher bursting measure gets better (i.e. the synchronization degree increases) via LTP of synaptic strengths, while bad burst synchronization with lower bursting measure gets worse (i.e. the synchronization degree decreases) via LTD. As a result, a step-like rapid transition to SBS occurs by changing $D$, in contrast to the relatively smooth transition in the absence of STDP.
In the presence of additive STDP, we also investigate the effect of network architecture on the SBS for a fixed $D$ by varying the symmetric attachment degree $l^*$ and the asymmetry parameter $\Delta l$ (tuning the asymmetrical attachment of new nodes with different in- and out-degrees) ($l_{in} = l^* + \Delta l$ and $l_{out}=l^*-\Delta l$; $l^* = 10$). Similar to the above case of the symmetric attachment with $l^*=10$, Matthew effects also occur by changing $l^*$ and $\Delta l$ (i.e., step-like rapid transitions to SBS take place, in contrast to the case without STDP).
Moreover, for the symmetric attachment with $l^*=10$, emergences of LTP and LTD of synaptic strengths are intensively studied through our own microscopic methods based on both the distributions
of time delays $\{ \Delta t_{ij} \}$ between the pre- and the post-synaptic bursting onset times and the pair-correlations between the pre- and the post-synaptic IIBRs (instantaneous individual burst rates).
To the best of our knowledge, there were no microscopic studies of this type in previous works on STDP. Hence, via these microscopic investigations, we also obtain another following main results, in addition to the Matthew
effect. We can clearly understand how microscopic distributions for $\{ \Delta t_{ij} \}$ contribute to the population-averaged synaptic modification $\langle J_{ij} \rangle$, and microscopic correlations between synaptic
pairs are also found to be directly associated with appearance of LTP/LTD. Finally, we consider a multiplicative STDP (which depends on states) \cite{Tass1,Multi}. For the multiplicative case, a change in synaptic strengths scales linearly with the distance to the higher and the lower bounds of synaptic strengths, and hence the bounds for the synaptic strength become ``soft,'' in contrast to the hard bounds for the additive case. The effects of multiplicative STDP on SBS for $l^*=10$ are investigated and discussed in comparison with the case of additive STDP. For this case of multiplicative STDP, a Matthew effect is also found to occur, as in the case of additive STDP. However, some quantitative differences arise, due to the effect of soft bounds. Consequently, a relatively less rapid transition occurs near both ends in comparison to the additive case, and the degrees of SBS in most plateau-like top region (corresponding to most cases of LTP) also become a little larger than those in the additive case.

This paper is organized as follows. In Sec.~\ref{sec:SFN}, we describe an excitatory Barab\'{a}si-Albert SFN of subthreshold Izhikevich neurons, and the governing
equations for the population dynamics are given. Then, in Sec.~\ref{sec:STDP} we investigate the effects of STDP on SBS for both cases of the additive and the multiplicative STDP.
Finally, in Sec.~\ref{sec:SUM} a summary is given.

\section{Excitatory Scale-Free Network of Subthreshold Neurons with Synaptic Plasticity}
\label{sec:SFN}
Synaptic connectivity in neural circuits has been found to have complex topology which is neither regular nor completely random \cite{Sporns,Buz2,CN1,CN2,CN3,CN4,CN5,CN6,CN7}.
Particularly, brain networks have been found to exhibit power-law degree distributions (i.e., scale-free property) in the rat hippocampal networks \cite{SF1,SF2,SF3,SF4} and the human cortical functional network \cite{SF5}. Moreover, robustness against simulated lesions of mammalian cortical anatomical networks \cite{SF6,SF7,SF8,SF9,SF10,SF11} has also been found to be most similar to that of an SFN \cite{SF12}.
Many recent works on various subjects of neurodynamics (e.g., coupling-induced burst synchronization, delay-induced burst synchronization, and suppression of burst synchronization) have been done in SFNs with a few percent of
hub neurons with an exceptionally large number of connections \cite{BSsync11,BSsync12,BSsync14,BSsync15,BSsync19,BSsync26}.

We consider an excitatory SFN composed of $N$ subthreshold neurons equidistantly placed on a one-dimensional ring of radius $N/ 2 \pi$. We employ a directed Barab\'{a}si-Albert SFN model (i.e. growth and preferential directed attachment) \cite{BA1,BA2}. At each discrete time $t,$ a new node is added, and it has $l_{in}$ incoming (afferent) edges and $l_{out}$ outgoing (efferent) edges via preferential attachments with $l_{in}$ (pre-existing) source nodes and $l_{out}$ (pre-existing) target nodes, respectively. The (pre-existing) source and target nodes $i$ (which are connected to the new node) are preferentially chosen depending on their out-degrees $d_i^{(out)}$ and in-degrees $d_i^{(in)}$ according to the attachment probabilities $\Pi_{source}(d_i^{(out)})$ and $\Pi_{target}(d_i^{(in)})$, respectively:
\begin{equation}
\begin{array}{l}
\Pi_{source}(d_i^{(out)})=\frac{d_i^{(out)}}{\sum_{j=1}^{N_{t -1}}d_j^{(out)}}\;\; \textrm{and} \\
\;\; \Pi_{target}(d_i^{(in)})=\frac{d_i^{(in)}}{\sum_{j=1}^{N_{t -1}}d_j^{(in)}},
\end{array}
\label{eq:AP}
\end{equation}
where $N_{t-1}$ is the number of nodes at the time step $t-1$.
Hereafter, the cases of $l_{in} = l_{out} (\equiv l^*)$  and $l_{in} \neq l_{out}$ will be referred to as symmetric and asymmetric preferential attachments, respectively.
For generation of an SFN with $N$ nodes, we start with the initial network at $t=0$, consisting of $N_0=50$ nodes where the node 1 is connected bidirectionally to all the other nodes, but the remaining nodes (except the node 1) are sparsely and randomly connected with a low probability $p=0.1$. Then, the processes of growth and preferential attachment are repeated until the total number of nodes becomes $N$. For our initial network, the node 1 will be grown as the head hub with the highest degree. As elements in the SFN, we choose the Izhikevich neuron model which combines the biological plausibility of the Hodgkin-Huxley-type models and the computational efficiency
of the integrate-and-fire model \cite{Izhi1,Izhi2}.

\begin{table}
\caption{Parameter values used in our computations; units of the potential and the time are mV and msec, respectively.}
\label{tab:Parm}
\begin{ruledtabular}
\begin{tabular}{llllll}
(1) & \multicolumn{5}{l}{Single Izhikevich Neurons \cite{Izhi1,Izhi2}} \\
&  $a=0.02$ & $b=0.2$ & $c=-65$ & $d=8$ & $v_p=30$ \\
\hline
(2) & \multicolumn{5}{l}{External Stimulus to Izhikevich Neurons} \\
& \multicolumn{2}{l}{$I_{DC,i} \in [3.55, 3.65]$} & \multicolumn{3}{l}{$D$: Varying} \\
\hline
(3) & \multicolumn{5}{l}{Excitatory Synapse Mediated by The AMPA} \\
 & \multicolumn{5}{l}{ Neurotransmitter \cite{AMPA}} \\
& $\tau_l=1$ & $\tau_r=0.5$ & $\tau_d=2$ & \multicolumn{2}{l}{$V_{syn}=0$} \\
\hline
(4) & \multicolumn{5}{l}{Synaptic Connections between Neurons in The} \\
& \multicolumn{5}{l}{Barab\'{a}si-Albert SFN} \\
& \multicolumn{5}{l}{$l^*$: Varying (symmetric preferential attachment)} \\
& \multicolumn{5}{l}{$\Delta l$: Varying (asymmetric preferential attachment)} \\
\hline
(5) & \multicolumn{5}{l}{Hebbian STDP Rule} \\
& $A_{+} = 1.0$ & $A_{-} = 0.6$ & $\tau_{+} = 15$ & $\tau_{-} = 30$ & \\
& $\delta = 0.005$ & \multicolumn{4}{l}{$J_{ij} \in [0.0001, 5.0]$} \\
\end{tabular}
\end{ruledtabular}
\end{table}

The following equations (\ref{eq:PD1})-(\ref{eq:PD6}) govern the population dynamics in the SFN:
\begin{eqnarray}
\frac{dv_i}{dt} &=& f(v_i) - u_i + I_{DC,i} +D \xi_{i} -I_{syn,i}, \label{eq:PD1} \\
\frac{du_i}{dt} &=& a~ (b v_i - u_i),  \;\;\; i=1, \cdots, N, \label{eq:PD2}
\end{eqnarray}
with the auxiliary after-spike resetting:
\begin{equation}
{\rm if~} v_i \geq v_p,~ {\rm then~} v_i \leftarrow c~ {\rm and~} u_i \leftarrow u_i + d, \label{eq:PD3}
\end{equation}
where
\begin{eqnarray}
f(v) &=& 0.04 v^2 + 5 v + 140,  \label{eq:PD4} \\
I_{syn,i} &=& \frac{1}{d_i^{(in)}} \sum_{j=1 (j \ne i)}^N J_{ij}~w_{ij}~s_j(t)~ (v_i - V_{syn}), \label{eq:PD5}\\
s_j(t) &=& \sum_{f=1}^{F_j} E(t-t_f^{(j)}-\tau_l);\nonumber \\ 
E(t) &=& \frac{1}{\tau_d - \tau_r} (e^{-t/\tau_d} - e^{-t/\tau_r}) \Theta(t). \label{eq:PD6}
\end{eqnarray}
Here, $v_i(t)$ and $u_i(t)$ are the state variables of the $i$th neuron at a time $t$ which represent the membrane potential and the recovery current, respectively. This membrane
potential and the recovery variable, $v_i(t)$ and $u_i(t)$, are reset according to Eq.~(\ref{eq:PD3}) when $v_i(t)$ reaches its cutoff value $v_p$. The parameter values used in our
computations are listed in Table \ref{tab:Parm}. More details on the Izhikevich neuron model, the external stimulus to each Izhikevich neuron, the synaptic currents and plasticity,
and the numerical method for integration of the governing equations are given in the following subsections.

\subsection{Izhikevich Neuron Model}
\label{subsec:Izhi}
The Izhikevich model matches neuronal dynamics by tuning the parameters $(a, b, c, d)$ instead of matching neuronal electrophysiology, unlike the Hodgkin-Huxley-type conductance-based
models \cite{Izhi1,Izhi2}. The parameters $a$, $b$, $c$, and $d$ are related to the time scale of the recovery variable $u$, the sensitivity of $u$ to the subthreshold fluctuations of $v$, and the after-spike reset
values of $v$ and $u$, respectively. Depending on the values of these parameters, the Izhikevich neuron model may exhibit 20 of the most prominent neuro-computational features of cortical
neurons, as in the Hodgkin-Huxley-type models. Here, we use the parameter values for the regular-spiking (RS) neurons, which are listed in the 1st item of Table \ref{tab:Parm}.

\subsection{External Stimulus to Each Izhikevich Neuron}
\label{subsec:Sti}
Each Izhikevich RS neuron is stimulated by both a DC current $I_{DC,i}$ and an independent Gaussian white noise $\xi_i$ [see the 3rd and the 4th terms in Eq.~(\ref{eq:PD1})].
The Gaussian white noise satisfies $\langle \xi_i(t) \rangle =0$ and $\langle \xi_i(t)~\xi_j(t') \rangle = \delta_{ij}~\delta(t-t')$, where $\langle\cdots\rangle$ denotes an ensemble
average. Here, the intensity of the Gaussian noise $\xi$ is controlled by the parameter $D$. For $D=0$, the Izhikevich RS neurons exhibit the type-II excitability. A type-II neuron
exhibits a jump from a resting state to a spiking state through a subcritical Hopf bifurcation when passing a threshold by absorbing an unstable limit cycle born via fold limit cycle
bifurcation, and hence the firing frequency begins from a non-zero value \cite{Izhi,Ex1}. Throughout the paper, we consider a subthreshold case (where only noise-induced firings occur)
such that the value of $I_{DC,i}$ is chosen via uniform random sampling in the range of [3.55, 3.65], as shown in the 2nd item of Table \ref{tab:Parm}.

\subsection{Synaptic Currents and Plasticity}
\label{subsec:Syn}
The 5th term in Eq.~(\ref{eq:PD1}) denotes the synaptic couplings of Izhikevich neurons. $I_{syn,i}$ of Eq.~(\ref{eq:PD5}) represents the synaptic current injected into the $i$th neuron,
and $V_{syn}$ is the synaptic reversal potential. The synaptic connectivity is given by the connection weight matrix $W$ (=$\{ w_{ij} \}$) where $w_{ij}=1$ if the neuron $j$ is presynaptic
to the neuron $i$; otherwise, $w_{ij}=0$. Here, the synaptic connection is modeled in terms of the directed Barab\'{a}si-Albert SFN. Then, the in-degree of the $i$th neuron, $d_{i}^{(in)}$ (i.e.,
the number of synaptic inputs to the neuron $i$) is given by $d_{i}^{(in)} = \sum_{j=1(\ne i)}^N w_{ij}$.

The fraction of open synaptic ion channels at time $t$ is denoted by $s(t)$. The time course of $s_j(t)$ for the $j$th neuron is given by a sum of delayed double-exponential functions
$E(t-t_f^{(j)}-\tau_l)$ [see Eq.~(\ref{eq:PD6})], where $\tau_l$ is the synaptic delay, and $t_f^{(j)}$ and $F_j$ are the $f$th spiking time and the total number of spikes of the $j$th neuron
(which occur until time $t$), respectively. Here, $E(t)$ [which corresponds to contribution of a pre-synaptic spike occurring at time $0$ to $s(t)$ in the absence of synaptic delay] is controlled
by the two synaptic time constants: synaptic rise time $\tau_r$ and decay time $\tau_d$, and $\Theta(t)$ is the Heaviside step function: $\Theta(t)=1$ for $t \geq 0$ and 0 for $t <0$. For the
excitatory AMPA synapse, the values of $\tau_l$, $\tau_r$, $\tau_d$, and $V_{syn}$ are listed in the 3rd item of Table \ref{tab:Parm} \cite{AMPA}.

\begin{figure}
\includegraphics[width=0.7\columnwidth]{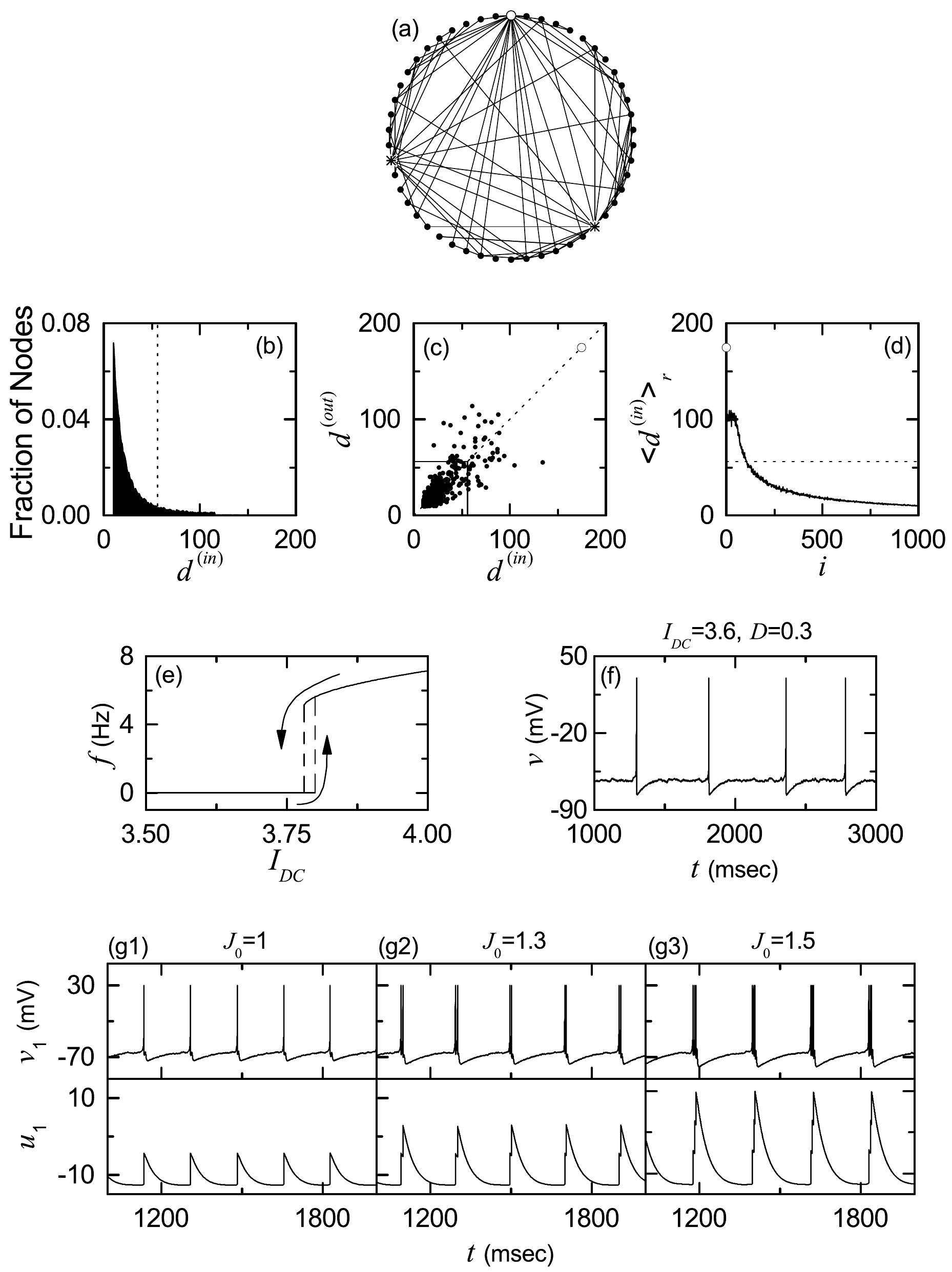}
\caption{SFN for the case of symmetrical attachment with $l_{in}=l_{out}=l^*=10$ when $N=10^3$. (a) Schematic diagram of an inhomogeneous SFN with 50 nodes equidistantly placed on a ring.
(b) Histogram for fraction of nodes versus the in-degree $d^{(in)}$.
(c) Plot of the out-degree $d^{(out)}$ versus the in-degree $d^{(in)}$. (d) Plot of the in-degree $\langle d^{(in)} \rangle_r$ versus the neuron index $i$. In c) and (d) the head hub is represented by the open circle.
Single Izhikevich RS neuron exhibiting type-II excitability. (e) Plot of the mean firing rate $f$ versus $I_{DC}$ for $D=0$. (f) Time series of the membrane potential $v(t)$ for $I_{DC}=3.6$ and $D=0.3$.
Coupling-induced transition from noise-induced spikings to noise-induced burstings for $D=0.3$ in the directed SFN of $N (=10^3)$ excitatory subthreshold Izhikevich RS neurons for the case of symmetrical attachment
with $l^*=10$. (g1)-(g3) Time series of the membrane potential $v_1(t)$ and the recovery variable $u_1(t)$ of the first neuron, where the mean values $J_0$ of synaptic coupling strengths $\{ J_{ij} \}$ are (g1) 1.0,
(g2) 1.3, and (g3) 1.5.
}
\label{fig:SN}
\end{figure}

The coupling strength of the synapse from the $j$th pre-synaptic bursting neuron to the $i$th post-synaptic bursting neuron is $J_{ij}$.
The values of $J_{ij}$ are obtained from the Gaussian distribution with the mean $J_0$ and the standard deviation $\sigma_0~(=0.02)$.
As $J_0$ passes a threshold, subthreshold Izhikevich RS neurons exhibit noise-induced burstings, which will be discussed in Fig.~\ref{fig:SN}. We are interested in SBS between
these noise-induced burstings.

Here, we consider a Hebbian STDP for the synaptic strengths $\{ J_{ij} \}$ and investigate effects of STDP on SBS. Initial synaptic strengths are normally distributed with
the mean $J_0~(=2.5)$ and the standard deviation $\sigma_0~(=0.02)$. With increasing time $t$, the synaptic strength for each synapse is updated with an additive nearest-burst pair-based STDP rule \cite{Tass1,SS}:
\begin{equation}
  J_{ij} \rightarrow J_{ij} + \delta~ \Delta J_{ij}(\Delta t_{ij}),
\label{eq:ASTDP}
\end{equation}
where $\delta$ $(=0.005)$ is the update rate and $\Delta J_{ij}$ is the synaptic modification depending on the relative time difference $\Delta t_{ij}$ $(=t_i^{(post)} - t_j^{(pre)})$
between the nearest burst onset times of the post-synaptic bursting neuron $i$ and the pre-synaptic bursting neuron $j$.
The synaptic modification $\Delta J_{ij}$ in Eq.~(\ref{eq:ASTDP}) for the case of burst synchronization is in contrast to the case of spike synchronization where $\Delta J_{ij}$  changes depending on the relative time difference between the nearest spike times of the post-synaptic and the pre-synaptic spiking neurons \cite{SSS}.  For a mixed case where neurons exhibit spikes and bursts, one can apply $\Delta J_{ij}$
in Eq.~(\ref{eq:ASTDP}) by treating each spike time as a burst onset time, because a spike may be regarded as a burst composed of only one spike.
To avoid unbounded growth, negative conductances (i.e. negative coupling strength), and elimination of synapses (i.e. $J_{ij}=0$), we set a range with the upper and the lower bounds: $J_{ij} \in [0.0001, 5.0]$.
We use an asymmetric time window for the synaptic modification $\Delta J_{ij}(\Delta t_{ij})$ \cite{STDP1}:
\begin{equation}
  \Delta J_{ij} = \left\{ \begin{array}{l} A_{+}~  e^{-\Delta t_{ij} / \tau_{+}} ~{\rm for}~ \Delta t_{ij} > 0\\
  - A_{-}~ e^{\Delta t_{ij} / \tau_{-}} ~{\rm for}~ \Delta t_{ij} < 0\end{array} \right. ,
\label{eq:TW}
\end{equation}
where $A_+=1.0$, $A_-=0.6$, $\tau_+=15$ msec, $\tau_-=30$ msec (these values are also given in the 5th item of Table \ref{tab:Parm}), and $\Delta J_{ij}(\Delta t_{ij}=0) = 0$.

\subsection{Numerical Method for Integration}
\label{subsec:NM}
Numerical integration of stochastic differential Eqs.~(\ref{eq:PD1})-(\ref{eq:PD6}) with a Hebbian STDP rule of Eqs.~(\ref{eq:ASTDP}) and (\ref{eq:TW})
is done by employing the Heun method \cite{SDE} with the time step $\Delta t=0.01$ msec. For each
realization of the stochastic process, we choose  random initial points $[v_i(0),u_i(0)]$ for the $i$th $(i=1,\dots, N)$ neuron with uniform probability in the range of
$v_i(0) \in (-50,-45)$ and $u_i(0) \in (10,15)$.

\section{Effects of STDP on the Stochastic Burst Synchronization}
\label{sec:STDP}
We consider a directed Barab\'{a}si-Albert SFN model with growth and preferential directed attachment \cite{BA1,BA2}. For reference, an inhomogeneous SFN (with 50 nodes equidistantly placed on a ring) is schematically depicted in Fig.~\ref{fig:SN}(a). There are a few of super-connected hubs with higher degrees, along with the majority of peripheral nodes with lower degrees. The head hub with the highest degree is denoted by the open circle, and two other secondary hubs are represented by the stars. We note that long-range connections (for global communication between distant nodes) emerge from these hubs.

Figures \ref{fig:SN}(b)-\ref{fig:SN}(d) show the degree distributions for the case of symmetric attachment with $l_{in}=l_{out}=l^*=10$ in the directed Barab\'{a}si-Albert SFN. The histogram for fraction of nodes
versus the in-degree $d^{(in)}$ is shown in Fig.~\ref{fig:SN}(b); this histogram is obtained through 30 realizations, and the bin size is 1. This in-degree distribution exhibits a power-law decay $P(d^{(in)})
\sim {d^{(in)}}^{-\gamma}$ with the exponent $\gamma=3$ \cite{BA1,BA2,Kim3}.
Hence, the majority of peripheral nodes have their degrees near the peak at $d^{(in)} = 10$, while the minority of hubs have their degrees in the long-tail part. Based on the degree distribution (showing a power-law
decay), we classify the nodes into the hub group (composed of the head hub with the highest degree and the secondary hubs with higher degrees) and the peripheral group (consisting of a majority of peripheral nodes with
lower degrees) in the following way \cite{Kim3,Kim4}. We choose an appropriate threshold $d_{th}^{(in)}$ separating the hub and the peripheral groups in the distribution of in-degrees $d^{(in)}$ in Fig.~\ref{fig:SN}(b).
For convenience, when the fraction of nodes is smaller than $0.2~\%$, such nodes are regarded as hubs. To this end, the threshold is chosen as $d_{th}^{(in)}=56$ [denoted by the vertical dotted line in Fig.~\ref{fig:SN}(b)]
whose fraction of nodes is 0.002 (i.e., $0.2~\%$). Figure \ref{fig:SN}(c) shows a plot of the out-degree $d^{(out)}$ versus the in-degree $d^{(in)}$. The in- and out-degrees are distributed nearly symmetrically around the
diagonal. Hence, we choose the threshold $d_{th}^{(out)}$ for the out-degree as $d_{th}^{(out)} =56$, which is the same as $d_{th}^{(in)}$. For visualization, the peripheral group is enclosed by a rectangle (determined by
both thresholds $d_{th}^{(in)}$ and $d_{th}^{(out)}$). The hub group (outside the rectangle) consists of 87 nodes (i.e., $8.7 \%$ of the total number $N~(=10^3)$ of neurons), where the node 1 (denoted by the open circle)
corresponds to the head hub with the highest degree and the other ones are secondary hubs. This type of degree distribution is a ``comet-shaped'' one; the peripheral and the hub groups correspond to the coma (surrounding
the nucleus) and the tail of the comet, respectively. Moreover, to find out which group (hub or peripheral) the neuron $i$ ($i=1, \dots, 1000)$ belongs to, we get a plot of the in-degree $d^{(in)}$ versus the neuron index
$i$ in Fig.~\ref{fig:SN}(d); $\langle \cdots \rangle_r$ denotes an average over 30 realizations. Here, nodes with smaller (larger) $i$ appear in the early (late) stage of the network evolution. The horizontal line represents
the threshold $(d^{(in)}=56)$ separating the hub and the peripheral neurons. Neurons with smaller $i$ are hubs, while those with larger $i$ are peripheral neurons

As elements in the SFN, we consider the Izhikevich RS neuron model \cite{Izhi1,Izhi2}. In the absence of noise ($D=0$), a single Izhikevich RS neuron exhibits a jump from a resting state to a spiking state via subcritical Hopf bifurcation at a higher threshold $I_{DC,h} (\simeq 3.80)$ by absorbing an unstable limit cycle born through a fold limit cycle bifurcation for a lower threshold $I_{DC,l} (\simeq 3.78)$ \cite{Kim1}. For this case, a plot of the mean firing rate $f$ versus the external DC current $I_{DC}$ is shown in Fig.~\ref{fig:SN}(e); each $f$ is obtained via an average for $10^5$ msec after a transient time of $10^3$ msec.
The Izhikevich RS neuron exhibits type-II excitability because it begins to fire with a non-zero frequency. As an example, we consider a subthreshold case of $I_{DC}=3.6$ in the presence of noise with $D=0.3$. This subthreshold Izhikevich RS neuron (which cannot fire spontaneously without noise) exhibits noise-induced spikings, as shown in Fig.~\ref{fig:SN}(f) for a time series of the membrane potential $v$.
Our SFN consists of $N~(=10^3)$ excitatory subthreshold Izhikevich RS neurons for the case of symmetrical attachment with $l_{in}=l_{out}=l^*=10$. The value of $I_{DC,i}$ for the $i$th neuron is chosen via uniform random sampling in the range of [3.55, 3.65]. The values of synaptic coupling strengths $J_{ij}$ between synaptic pairs are obtained from the Gaussian distribution with the mean $J_0$ and the standard deviation $\sigma_0~(=0.02)$, and they are static (i.e. absence of STDP). As shown in Figs.~\ref{fig:SN}(g1)-\ref{fig:SN}(g3), with increasing $J_0$ for a fixed value of $D=0.3$, coupling-induced transition from noise-induced spikings to noise-induced burstings occurs when passing a threshold $J^*_0 \simeq 1.207$ \cite{Kim1}. Figure \ref{fig:SN}(g1) shows the time series of the membrane potential $v_1$ and the recovery variable $u_1$ of the first neuron (in the population) for $J_0=1.0$.
The fast membrane potential $v_1$ exhibits a spiking or quiescent state depending on the slow recovery variable $u_1$ which provides a negative feedback to $v_1$ and can be regarded as an adaptation parameter \cite{Izhi,Burst3,Burstsync2}. For the case of $J_0=1.0$ (which is less than the critical value $J^*_0$), spiking $v_1$ pushes $u_1$ outside the spiking area. Then, $u_1$ makes a slow decay into the quiescent area
[see Fig.~\ref{fig:SN}(g1)], which leads to termination of spiking. The quiescent $v_1$ pushes $u_1$ outside the quiescent area, and then $u_1$ revisits the spiking area, which results in spiking of $v_1$.
Via repetition of this process, noise-induced spikings appear successively in $v_1$ for $J_0=1.0$. However, when passing a threshold $J^*_0$, the coherent synaptic input to the first neuron becomes so strong that the first spike in $v_1$ cannot push $u_1$ outside the spiking area. As an example, see the case of $J_0=1.3$ in Fig.~\ref{fig:SN}(g2). In this case, after the 1st spike in $v_1$, $u_1$ at first decreases only a little, and then it increases abruptly. Unlike the case of $J_0=1.0$, after the 1st spike, $u_1$ remains inside the spiking area, and hence a second spike appears in $v_1$. After the 2nd spike, $u_1$ is pushed away from the spiking area and slowly
decays into the quiescent area, which leads to termination of repetitive spikings. Consequently, noise-induced burstings, composed of two spikes (doublets), appear in $v_1$ for $J_0=1.3$. With further increasing $J_0$, the coherent synaptic input becomes stronger, and hence the number of spikes in a noise-induced bursting increases [e.g., see the noise-induced triplets in Fig.~\ref{fig:SN}(g3) for $J_0=1.5$].

\subsection{SBS in The Absence of STDP}
\label{subsec:NSTDP}
\begin{figure}
\includegraphics[width=\columnwidth]{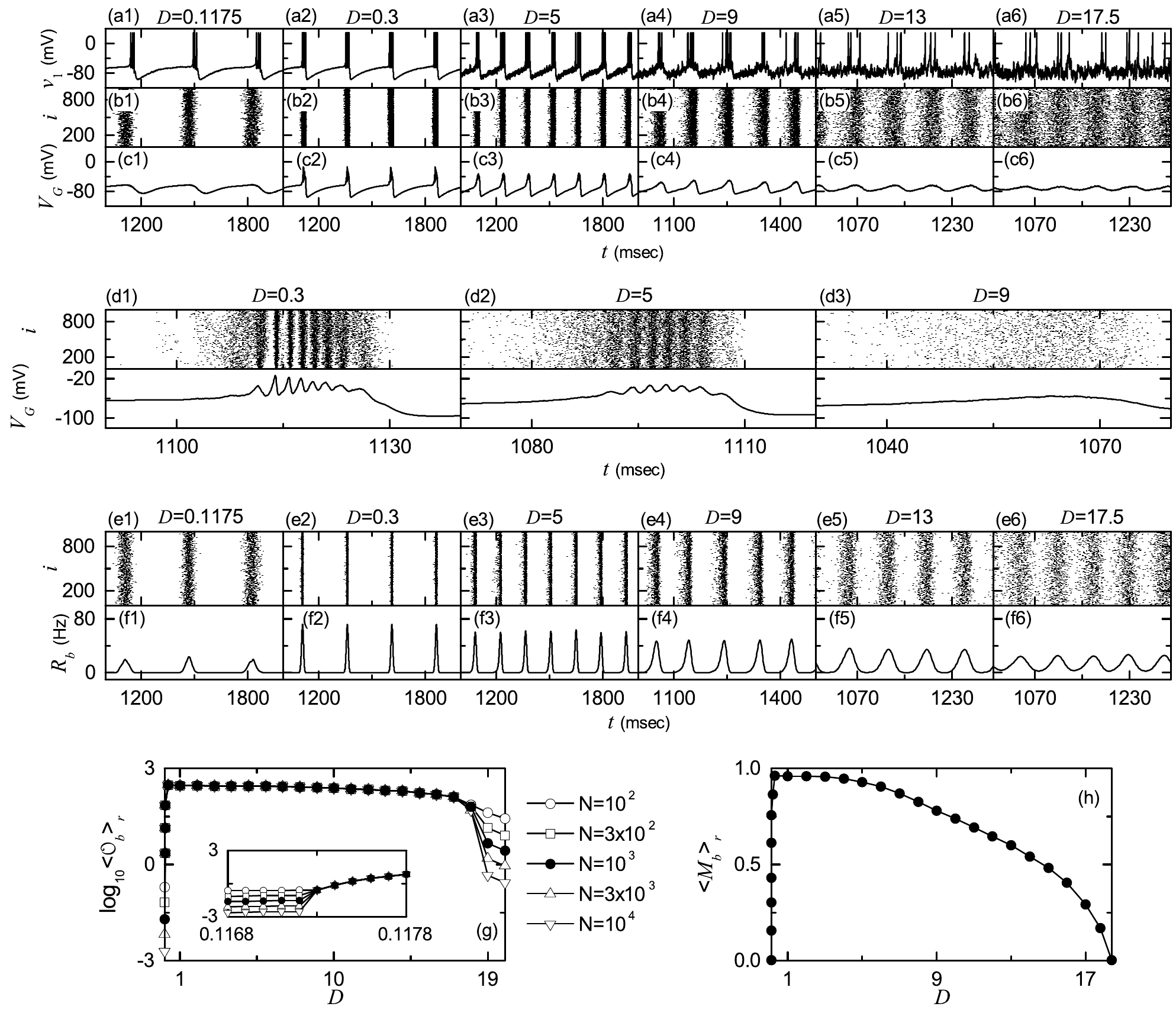}
\caption{SBS in the absence of STDP for the case of symmetrical attachment with $l^*=10$; $N=10^3$ except for the case in (g). Time series of the membrane potential $v_1(t)$ of the 1st neuron in (a1)-(a6),
raster plots of spikes in (b1)-(b6), and time series of the global potential $V_G(t)$ in (c1)-(c6) for various values of $D$. Raster plots of spikes and time series of the global potential $V_G(t)$ for
a single magnified burst for $D=$ (d1) 0.3, (d2) 5, and (d3) 9. Raster plots of burst onset times in (e1)-(e6) and IPBR kernel estimates $R_b(t)$ in (f1)-(f6) for various values of $D$. (g) Plots of the
thermodynamic bursting order parameter $\log_{10} \langle {\cal{O}}_b \rangle_r$ versus $D$. (h) Plot of the statistical-mechanical bursting measure $\langle M_b \rangle$ versus $D$.
}
\label{fig:SBS1}
\end{figure}

First, we are concerned about the SBS in the absence of STDP for the case of symmetric attachment with $l_{in}= l_{out}=l^*=10$
in the SFN of $N$ excitatory subthreshold Izhikevich neurons. The coupling strengths $\{ J_{ij} \}$ are static, and their values are chosen from the Gaussian distribution where the mean
$J_0$ is 2.5 and the standard deviation $\sigma_0$ is 0.02. We investigate emergence of SBS (i.e., population synchronization between noise-induced burstings) by varying the noise intensity $D$.
Figures \ref{fig:SBS1}(a1)-\ref{fig:SBS1}(a6) show the time series of $v_1$ of the 1st neuron for various values of $D$.
For sufficiently small $D$ [which is less than the lower threshold $D^*_l ~(\simeq 0.1173)]$, individual neurons exhibit sparse noise-induced spikings because there are no coherent synaptic inputs.
When passing $D^*_l$, noise-induced (``regular'') burstings appear due to strong coherent synaptic inputs
(resulting from a constructive role of noise to stimulate coherence between noise-induced firings) [e.g., see Figs.~\ref{fig:SBS1}(a1)-\ref{fig:SBS1}(a2)].
However, with further increase in $D,$ some irregularities begin to occur in both the number of spikes and the interspike intervals within the noise-induced burstings due to a destructive role of noise to spoil the
population coherence, as shown in Figs.~\ref{fig:SBS1}(a3)-\ref{fig:SBS1}(a6). Eventually, when passing the higher threshold $D^*_h ~(\simeq 18.4)$, such irregularities become so intensified that individual
neurons exhibit irregular mixed (noise-induced) burstings and spikings.

Population synchronization may be well visualized in the raster plot of neural spikes which is a collection of spike trains of individual neurons. Raster plots of spikes are shown in Figs.~\ref{fig:SBS1}(b1)-\ref{fig:SBS1}(b6)
for various values of $D$. Such raster plots of spikes are fundamental data in experimental neuroscience. As a collective quantity showing population behaviors, we also consider the population-averaged membrane potential $V_G$ (corresponding to the global potential):
\begin{equation}
 V_G (t) = \frac {1} {N} \sum_{i=1}^{N} v_i(t).
\label{eq:GP}
\end{equation}
Global potentials $V_G$ for various values of $D$ are shown in Figs.~\ref{fig:SBS1}(c1)-\ref{fig:SBS1}(c6).
For the synchronous case, ``stripes'' (composed of spikes and indicating population synchronization) are found to be formed in the raster plot of spikes, and an oscillating global potential $V_G$ appears
[see Figs.~\ref{fig:SBS1}(b1)-\ref{fig:SBS1}(b6) and Figs.~\ref{fig:SBS1}(c1)-\ref{fig:SBS1}(c6)]. On the other hand, in the desynchronized case for $D < D^*_l$ or $D> D^*_h$, spikes are completely scattered in the raster plot of spikes, and $V_G$ is nearly stationary.
For a clear view, magnifications of a single bursting band and $V_G$ are given in Figs.~\ref{fig:SBS1}(d1)-\ref{fig:SBS1}(d3) for $D=0.3$, 5, and 9, respectively.

As mentioned in Sec.~\ref{sec:INT}, bursting neurons exhibit two different types of synchronization due to the slow and the fast timescales of bursting activity. Burst synchronization (synchrony on the slow bursting timescale) refers to a temporal coherence between burst onset times (i.e., times at which burstings begin in bursting bands), while spike synchronization (synchrony on the fast spiking timescale) characterizes a temporal coherence between intraburst spikes fired by bursting neurons in their respective active phases \cite{Burstsync1,Burstsync2}. When both burst synchronization with the slow timescale and intraburst spike synchronization with the fast timescale occur, we call it as complete synchronization. For $D=0.3$, slow burst synchronization occurs, because bursting bands appear regularly in the raster plot [see Fig.~\ref{fig:SBS1}(b2)]. Furthermore, since each burst band is composed of intraburst spiking stripes [see Fig.~\ref{fig:SBS1}(d1)], fast intraburst spike synchronization also occurs. Consequently, complete synchronization (including both slow burst synchronization and fast intraburst spike synchronization) occurs for D=0.3. Hence, the global potential $V_G$ for $D=0.3$ exhibits a bursting activity like the individual membrane potentials (i.e., fast spikes appear on a slow wave) [see Fig.~\ref{fig:SBS1}(d1)].
However, as $D$ is increased, loss of spike synchronization occurs due to smearing of spiking stripes in each burst band.
As an example, see the case of $D=5$ where magnifications of a single burst band and $V_G$ are given in Fig.~\ref{fig:SBS1}(d2). Smearing of spiking stripes is well seen in the magnified burst band, and hence
the amplitudes of spikes on the slow wave in $V_G$ decrease. As $D$ is further increased and passes a (higher) threshold $D^{(s)}_h~(\simeq 7.7$), complete loss of spike synchronization occurs in each burst band
(i.e., a transition from complete synchronization to burst synchronization occurs). As a result, only burst synchronization (without spike synchronization) occurs, as shown in Fig.~\ref{fig:SBS1}(d3) for $D=9$. In this case, $V_G$ exhibits a slow-wave oscillation without fast spikes.
We also note that for small $D$ just above the lower threshold $D^*_l$ (e.g., see the case of $D=0.1175$), only burst synchronization occurs, as shown in Figs.~\ref{fig:SBS1}(b1) and \ref{fig:SBS1}(c1)
where no spiking stripes are formed in each burst band, and hence only a slow-wave oscillation appears in $V_G$. As $D$ is a little more increased and passes a (lower) threshold $D^{(s)}_l~(\simeq 0.1196)$, a transition from
burst synchronization to complete synchronization occurs. Consequently, burst synchronization emerges in the whole range of $D^*_l < D < D^*_h$, while complete synchronization (including both burst and spike synchronization)
appears in a sub-range of $D^{(s)}_l < D < D^{(s)}_h$.

Hereafter, we pay attention to only burst synchronization (i.e., population synchronization on the slow bursting timescale) without considering fast (intraburst) spike synchronization.
For more direct visualization of just bursting behaviors, we consider another raster plot of burst onset times (i.e., times at which burstings begin in bursting bands).
For convenience, we choose the 1st spike time in each bursting band as the burst onset time. In this way, the burst onset time (i.e., the 1st spike time) becomes a representative bursting time in each bursting band.
A collection of all trains of burst onset times of individual neurons forms a raster plot of burst onset times [e.g., see Figs.~\ref{fig:SBS1}(e1)-\ref{fig:SBS1}(e6)], which is in contrast to raster plots of spikes (i.e., collections of spike trains of individual neurons) where all intraburst spike times are considered [e.g., see Figs.~\ref{fig:SBS1}(b1)-\ref{fig:SBS1}(b6)]. The raster plot of burst onset times contains all essential information on the bursting behaviors.
Figures \ref{fig:SBS1}(e1)-\ref{fig:SBS1}(e6) show raster plots of burst onset times for various values of $D$.
To see emergence of burst synchronization, we employ an (experimentally-obtainable) instantaneous population burst rate (IPBR) which is often used as a collective quantity showing bursting behaviors.
This IPBR may be obtained from the raster plot of burst onset times \cite{Kim2,Kim4,Kim5}. To obtain a smooth IPBR, we employ the kernel density estimation (kernel smoother) \cite{Kernel}. Each burst onset time in the raster plot is convoluted (or blurred) with a kernel function $K_h(t)$ to obtain a smooth estimate of IPBR $R_b(t)$:
\begin{equation}
R_b(t) = \frac{1}{N} \sum_{i=1}^{N} \sum_{b=1}^{n_i} K_h (t-t_{b}^{(i)}),
\label{eq:IPSR}
\end{equation}
where $t_{b}^{(i)}$ is the $b$th burst onset time of the $i$th neuron, $n_i$ is the total number of burst onset times for the $i$th neuron, and we use a Gaussian
kernel function of band width $h$:
\begin{equation}
K_h (t) = \frac{1}{\sqrt{2\pi}h} e^{-t^2 / 2h^2}, ~~~~ -\infty < t < \infty \label{eq:Gaussian}
\end{equation}
Throughout the paper, the band width $h$ of $K_h(t)$ is 5 msec. Figures \ref{fig:SBS1}(f1)-\ref{fig:SBS1}(f6) show IPBR kernel estimates $R_b(t)$ for various values of $D$.
For the synchronous case, ``bursting stripes'' (composed of burst onset times and indicating burst synchronization) are formed in the raster plot of burst onset times [see Figs.~\ref{fig:SBS1}(e1)-\ref{fig:SBS1}(e6)], and the corresponding IPBR kernel estimates $R_b(t)$ exhibit oscillations, as shown in Figs.~\ref{fig:SBS1}(f1)-\ref{fig:SBS1}(f6). The bursting frequency $f_b$ [i.e., the oscillating frequency of $R_b(t)$] increases with increasing $D$.
(e.g., for $D=0.1175,$ $f_b \simeq 2.8$ Hz, while for $D=17.5,$ $f_b \simeq 16.7$ Hz). In contrast, in the desynchronized case for $D < D^*_l$ or $D > D^*_h$, burst onset times are completely scattered in the raster plot, and $R_b(t)$ is nearly stationary.

Recently, we introduced a realistic bursting order parameter, based on $R_b(t)$, for describing transition from desynchronization to burst synchronization \cite{Kim5}.
The mean square deviation of $R_b(t)$,
\begin{equation}
{\cal{O}}_b \equiv \overline{(R_b(t) - \overline{R_b(t)})^2},
 \label{eq:Order}
\end{equation}
plays the role of an order parameter ${\cal{O}}_b$; the overbar represents the time average. This bursting order parameter may be regarded as a thermodynamic measure because it
concerns just the macroscopic IPBR kernel estimate $R_b(t)$ without any consideration between $R_b(t)$ and microscopic individual burst onset times.
In the thermodynamic limit of $N \rightarrow \infty$, the bursting order parameter ${\cal{O}}_b$ approaches a non-zero (zero) limit value for the synchronized (desynchronized) state.
Hence, the bursting order parameter can determine synchronized and desynchronized states for the case of the burst synchronization.
Figure \ref{fig:SBS1}(g) shows plots of $\log_{10} \langle {\cal{O}}_b \rangle_r$ versus $D$.
In each realization, we discard the first time steps of a stochastic trajectory as transients for $10^3$ msec, and then we numerically compute ${\cal{O}}_b$ by following the stochastic trajectory
for $3 \times 10^4$ msec. Hereafter, $\langle \cdots \rangle_r$ denotes an average over 20 realizations.
For $D < D^*_l$ $(\simeq 0.1173$), desynchronized states exist because the bursting order parameter ${\cal{O}}_b$ tends to zero as
$N \rightarrow \infty$. As $D$ passes the lower threshold $D^*_l$, a transition to SBS occurs due to a constructive role of noise stimulating coherence
between noise-induced burstings of subthreshold neurons. However, for large $D > D^*_h$ $(\simeq 18.4)$ such synchronized states disappear (i.e., a transition to
desynchronization occurs when $D$ passes the higher threshold $D^*_h$) due to a destructive role of noise spoiling the SBS. In this way, SBS appears in an
intermediate range of $D^*_l < D < D^*_h$ through competition between the constructive and the destructive roles of noise.
For $D < D^*_l$ burst onset times are scattered without forming any stripes in the raster plot, and hence the IPBR kernel estimate $R_b(t)$ is nearly stationary. On the other hand, when passing $D^*_l$, synchronized states appear. As shown in Figs.~\ref{fig:SBS1}(e1) and \ref{fig:SBS1}(f1) for $D=0.1175,$ wide bursting stripes (indicating burst synchronization) appear successively in the raster plot of burst onset times, and the IPBR kernel estimate $R_b(t)$ exhibits an oscillatory behavior. With a little increase in $D,$ the degree of SBS is abruptly increased
because clearer narrowed bursting stripes appear in the raster plot (e.g., see the case of $D=0.3$). As a result, the amplitude of $R_b(t)$ also increases so rapidly.
However, with further increase in $D$, bursting stripes become smeared gradually, as shown in the cases of $D=5,$ 9, 13, and 17.5, and hence the amplitudes of $R_b (t)$ decreases in a slow way.
Eventually, when passing $D^*_h,$ desynchronization occurs due to overlap of smeared bursting stripes.

We characterize SBS by employing a statistical-mechanical bursting measure $M_b$ \cite{Kim5}. For the case of SBS, bursting stripes appear regularly in the raster plot of burst onset times.
The bursting measure $M^{(b)}_i$ of the $i$th bursting stripe is defined by the product of the occupation degree $O^{(b)}_i$ of burst onset times (denoting the density of the $i$th bursting stripe) and the
pacing degree $P^{(b)}_i$ of burst onset times (representing the smearing of the $i$th bursting stripe):
\begin{equation}
M^{(b)}_i = O^{(b)}_i \cdot P^{(b)}_i.
\label{eq:BMi}
\end{equation}
The occupation degree $O^{(b)}_i$ of burst onset times in the $i$th bursting stripe is given by the fraction of bursting neurons:
\begin{equation}
   O^{(b)}_i = \frac {N_i^{(b)}} {N},
\label{eq:OD}
\end{equation}
where $N_i^{(b)}$ is the number of bursting neurons in the $i$th bursting stripe.
For the case of full synchronization, all bursting neurons exhibit burstings in each bursting stripe in the raster plot of burst onset times, and hence the occupation degree $O_i^{(b)}$
of Eq.~(\ref{eq:OD}) in each bursting stripe becomes 1. On the other hand, in the case of partial synchronization, only some fraction of bursting neurons show burstings in each bursting stripe,
and hence the occupation degree $O_i^{(b)}$ becomes less than 1.
In our case of SBS, $O^{(b)}_i=1$, independently of $D$. For this case of full synchronization, $M^{(b)}_i = P^{(b)}_i$.
The pacing degree $P^{(b)}_i$ of burst onset times in the $i$th bursting stripe can be determined in a statistical-mechanical way by taking into account their contributions to the macroscopic IPBR kernel estimate $R_b(t)$.
Central maxima of $R_b(t)$ between neighboring left and right minima of $R_b(t)$ coincide with centers of bursting stripes in the raster plot. A global cycle starts from a left minimum of
$R_b(t)$, passes a maximum, and ends at a right minimum.
An instantaneous global phase $\Phi^{(b)}(t)$ of $R_b(t)$ was introduced via linear interpolation in the region forming a global cycle
(for details, refer to Eqs.~(14) and (15) in \cite{Kim5}).  Then, the contribution of the $k$th microscopic burst onset time in the $i$th bursting stripe occurring at the time $t_k^{(b)}$ to $R_b(t)$ is
given by $\cos \Phi^{(b)}_k$, where $\Phi^{(b)}_k$ is the global phase at the $k$th burst onset time [i.e., $\Phi^{(b)}_k \equiv \Phi^{(b)}(t_k^{(b)})$]. A microscopic burst onset time makes the most constructive (in-phase)
contribution to $R_b(t)$ when the corresponding global phase $\Phi^{(b)}_k$ is $2 \pi n$ ($n=0,1,2, \dots$), while it makes the most destructive (anti-phase) contribution to $R_b(t)$ when $\Phi^{(b)}_k$
is $2 \pi (n-1/2)$. By averaging the contributions of all microscopic burst onset times in the $i$th bursting stripe to $R_b(t)$, we obtain the pacing degree of burst onset times in the $i$th stripe:
\begin{equation}
 P^{(b)}_i ={ \frac {1} {B_i}} \sum_{k=1}^{B_i} \cos \Phi^{(b)}_k,
\label{eq:PACING}
\end{equation}
where $B_i$ is the total number of microscopic burst onset times in the $i$th stripe.
By averaging $P^{(b)}_i$ over a sufficiently large number $N_b$ of bursting stripes, we obtain the realistic statistical-mechanical bursting measure $M_b$, based on the IPBR kernel estimate $R_b(t)$:
\begin{equation}
M_b =  {\frac {1} {N_b}} \sum_{i=1}^{N_b} P^{(b)}_i.
\label{eq:BM}
\end{equation}
We follow $3 \times 10^3$ bursting stripes in each realization and get $\langle M_b \rangle_r$ via average over 20 realizations. Figure \ref{fig:SBS1}(h) shows a plot of $\langle M_b \rangle_r$
(denoted by open circles) versus $D$. When passing $D^*_l$ a rapid increase in $\langle M_b \rangle_r$ occurs, then $\langle M_b \rangle_r$ decreases slowly near the region of complete
synchronization (including both burst and spike synchronization) because spike synchronization is first destroyed, and finally $\langle M_b \rangle_r$ decreases in a relatively rapid way in a larger region of
(pure) burst synchronization.

We now fix the value of $D$ at $D=13$ where only the burst synchronization (without intraburst spike synchronization) occurs for the case of symmetric attachment with $l^*=10$ [see Figs.~\ref{fig:SBS1}(e5) and \ref{fig:SBS1}(f5)], and investigate the effect of scale-free connectivity on SBS by varying (1) the degree of symmetric attachment $l^*$ (i.e., $l_{in}=l_{out}  = l^*$) and (2) the asymmetry parameter $\Delta l$ of asymmetric attachment [i.e., $l_{in}= l^* + \Delta l$ and $l_{out}= l^* - \Delta l$ ($l^*=10$)].

\begin{figure}
\includegraphics[width=\columnwidth]{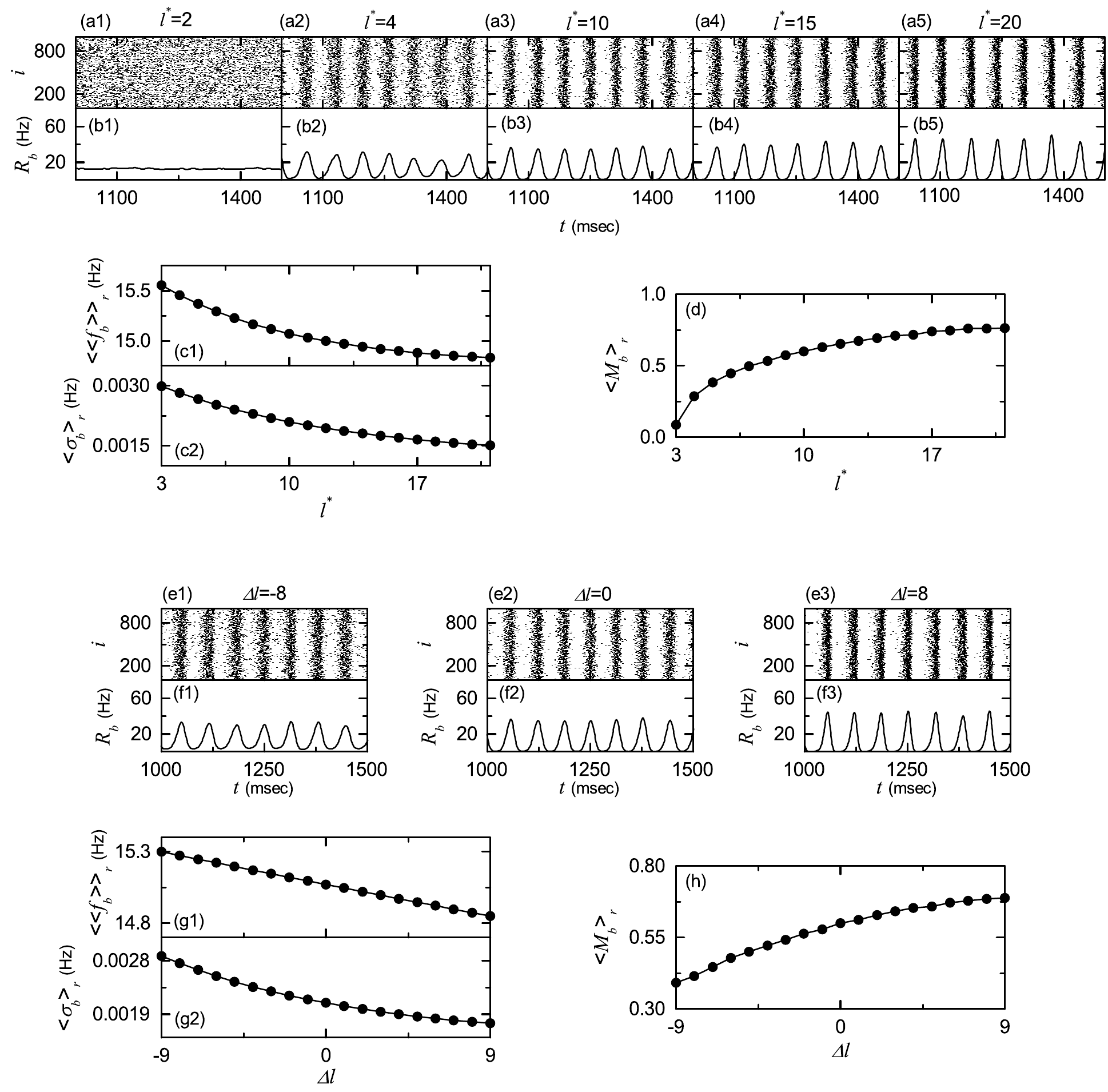}
\caption{Effect of network architecture on the SBS in the absence of STDP for $D=13$; $N=10^3$. Symmetric preferential attachment with $l_{in}=l_{out}=l^*$. Raster plots of burst onset times in (a1)-(a5)
and IPBR kernel estimates $R_b(t)$ in (b1)-(b5) for various values of $l^*$. Plots of (c1) population-averaged MBRs $\langle \langle f_b \rangle \rangle_r$ and
(c2) standard deviations $\langle \sigma_b \rangle_r$ from $\langle f_b \rangle$ versus $l^*$. (d) Plot of the statistical-mechanical bursting measure $\langle M_b \rangle_r$ versus $l^*$.
Asymmetric preferential attachment with $l_{in}= l^* + \Delta l$ and $l_{out} = l^* - \Delta l$ ($l^*=10$). Raster plots of burst onset times in (e1)-(e3) and IPBR kernel estimates $R_b(t)$
in (f1)-(f3) for various values of $\Delta l$. (g1) Plots of population-averaged MBRs $\langle \langle f_b \rangle \rangle_r$ and (g2) standard deviations $\langle \sigma_b \rangle_r$ from $\langle f_b \rangle$ versus $\Delta l$. (h) Plot of the statistical-mechanical bursting measure $\langle M_b \rangle_r$ versus $\Delta l$.
}
\label{fig:SBS2}
\end{figure}

As the first case of network architecture, we consider the case of symmetric attachment, and study  its effect on SBS by varying the degree $l^*$. Figures \ref{fig:SBS2}(a1)-\ref{fig:SBS2}(a5) show the raster plots of burst onset times for various values of $l^*$. Their corresponding IPBR kernel estimates $R_b(t)$ are also given in Figs.~\ref{fig:SBS2}(b1)-\ref{fig:SBS2}(b5). As $l^*$ is increased from $10$ (i.e., the case studied above), bursting stripes in the raster plots of burst onset times become clearer (e.g., see the cases of $l^*=15$ and 20), and hence the oscillating amplitudes of $R_b(t)$ become larger than that for the case of $l^*=10$. In this way, with increasing $l^*$ from 10, the degree of SBS becomes better. On the other hand, as $l^*$ is decreased from 10, bursting stripes become more smeared (e.g., see the case of $l^*=4$), which results in decrease in the oscillating amplitude of $R_b(t)$. Thus, with decreasing $l^*$ from 10, the degree of SBS becomes worse. Eventually, the population state becomes desynchronized for $l^*=2$, as shown in Figs.~\ref{fig:SBS2}(a1) and \ref{fig:SBS2}(b1) where burst onset times are completely scattered and $R_b(t)$ becomes nearly stationary.

Effects of $l^*$ on network topology were characterized in Refs.~\cite{Kim3,Kim4}, where the group properties of the SFN were studied in terms of the average path length $L_p$ and the betweenness centralization $B_c$
by varying $l^*$. The average path length $L_p$ (representing typical separation between two nodes in the network) is obtained via the average of the shortest path lengths of all nodal pairs (see Eq.~(A.17) in \cite{Kim4}), and it characterizes global efficiency of information transfer between distant nodes \cite{BA2}. With increasing $l^*$, $L_p$ decreases monotonically due to increase in the total number of inward and outward connections (see Fig.~11(c) in \cite{Kim4}). Next, we consider the betweenness centrality $B_i$ of the node $i$, denoting the fraction of all the shortest paths between any two other nodes that pass the node $i$
(see Eq.~(A.18) in \cite{Kim4}). The betweenness centrality $B_i$ characterizes the potentiality in controlling communication between other nodes in the rest of the network \cite{BETC1,BETC2}.
In our SFN, the head hub (i.e., node 1) has the maximum betweenness centrality $B_{max}$, and hence it has the largest load of communication traffic passing through it.
To examine how much the load of communication traffic is concentrated on the head hub, we get the group betweenness centralization $B_c$, denoting the degree to which the maximum betweenness centrality $B_{max}$ of the head hub exceeds the betweenness centralities of all the other nodes (see Eq.~(A.19) in \cite{Kim4}). Large $B_c$ implies that load of communication traffic is much concentrated on the head hub, and hence the head hub tends to become overloaded by the communication traffic passing through it. Consequently, it becomes difficult to obtain efficient communication between nodes due to destructive interference between many signals passing through the head hub \cite{BC3}. Decrease in $L_p$ with increasing $l^*$ leads to reduction in intermediate mediation of nodes controlling the communication in the whole network. Hence, as $l^*$ is increased, the total centrality $B_{tot}$, given by the sum of betweenness centralities $B_i$ of all nodes, is reduced. Particularly, with increasing $l^*$ the maximum betweenness $B_{max}$ of the head hub is much more reduced than betweenness centralities of any other nodes, which leads to decrease in differences between $B_{max}$ of the head hub and $B_i$ of other nodes. Consequently, with increasing $l^*$ the betweenness centralization $B_c$ decreases monotonically (see Fig.~11(e) in \cite{Kim4}).
In this way, as $l^*$ is increased, the average path length $L_p$ becomes smaller and the betweenness centralization $B_c$ also becomes smaller, due to increase in the total number of connections. Hence, typical separation between neurons (placed at nodes) becomes shorter, and load of communication traffic concentrated on the head neuron (placed at the head hub) also becomes smaller. Consequently, with increasing $l^*$, efficiency of global communication between neurons (i.e., global transfer of neural information between neurons via synaptic connections)  becomes better, which may lead to increase in the degree of SBS.

In addition to network topology, we also consider individual dynamics which vary depending on the synaptic inputs with the in-degree $d^{(in)}$ of Eq.~(\ref{eq:PD5}). As $l^*$ is increased, the average
in-degree $\langle d^{(in)} \rangle$ (=$ \frac {1} {N} \sum_{i=1}^{N} d^{(in)}_i$) increases, and hence average synaptic inputs to individual neurons become more coherent. Consequently, with increasing $l^*$, burstings of individual neurons become intensified (i.e., both the average number of spikes per burst and the average interburst interval increase), similar to the case of increasing $J_0$ in Figs.~\ref{fig:SN}(g1)-\ref{fig:SN}(g3).
Thus, as $l^*$ is increased, both the population-averaged mean bursting rate (MBR) $\langle \langle f_b \rangle \rangle_r$ and the standard deviation $\langle \sigma_b \rangle_r$
(for the distribution of MBRs $\{ f_b \}$) decrease (i.e., population-averaged individual dynamics become better) due to more coherent synaptic inputs (resulting from the increased $\langle d^{(in)} \rangle$), as shown in Fig.~\ref{fig:SBS2}(c1) and \ref{fig:SBS2}(c2), which may also result in increase in the degree of SBS.

Figure \ref{fig:SBS2}(d) shows a plot of the bursting measure $\langle M_b \rangle_r$ versus $l^*$. With increasing $l^*$ from 10, $\langle M_b \rangle_r$ increases due to both better individual dynamics
and better efficiency of global communication between nodes (resulting from the increased number of total connections). On the other hand, as $l^*$ is decreased from 10, both individual dynamics and effectiveness of
communication between nodes become worse (resulting from the decreased number of total connections), and hence $\langle M_b \rangle_r$ decreases.

As the second case of network architecture, we consider the case of asymmetric attachment; $l_{in}= l^* + \Delta l$ and $l_{out}= l^* - \Delta l$ ($l^*=10$). We note that for the case of asymmetric attachment, the total number of
inward and outward connections is fixed (i.e., $ l_{in} + l_{out} = 20$ =constant), in contrast to the case of symmetric attachment where with increasing $l^*$ the number of total connections increases. We investigate the effect
of asymmetric attachment on SBS by varying the asymmetry parameter $\Delta l$.

Figures \ref{fig:SBS2}(e1)-\ref{fig:SBS2}(e3) show the raster plots of burst onset times for $\Delta l=$ -8, 0, and 8, respectively. Their corresponding IPBR kernel estimates $R_b(t)$ are also given in Figs.~\ref{fig:SBS2}(f1)-\ref{fig:SBS2}(f3). As $\Delta l$ is increased from $0$, bursting stripes in the raster plots of burst onset times become clearer (e.g., see the cases of $\Delta l=8$), and hence the oscillating amplitudes of $R_b(t)$ become larger than that for the case of $\Delta l=0$. In this way, with increasing $\Delta l$ from 0, the degree of SBS becomes better. On the other hand, as $\Delta l$ is decreased from 0, bursting stripes become more smeared (e.g., see the case of $\Delta l = -8$), which leads to decrease in the oscillating amplitudes of $R_b(t)$.
Thus, as $\Delta l$ is decreased from 0, the degree of SBS becomes worse. For the present case of $l^*=10$, the minimum value of $\Delta l$ to be decreased is -9; in this case SBS persists.

As $| \Delta l |$ (the magnitude of $\Delta l$) is increased, both $L_p$ and $B_c$ increase symmetrically, independently of the sign of $\Delta l$, due to increased mismatching between the in- and the out-degrees
(see Figs.~13(c) and 13(d) in \cite{Kim4}). The values of $L_p$ and $B_c$ for both cases of different signs but the same magnitude (i.e., $\Delta l$ and $- \Delta l$) become the same because both inward and
outward connections are involved equally in computations of $L_p$ and $B_c$. As results of effects of $\Delta l$ on $L_p$ and $B_c$, with increasing $| \Delta l |$, efficiency of global communication between nodes becomes
worse, independently of the sign of $\Delta l$.
However, individual dynamics vary depending on the sign of $\Delta l$ due to different average in-degrees $\langle d^{(in)} \rangle$. As $\Delta l$ is increased (decreased) from 0, $\langle d^{(in)} \rangle$ increases (decreases), which leads to more (less) coherent synaptic inputs to individual neurons. Hence, with increasing (decreasing) $\Delta l$ from 0, both the population-averaged MBR $\langle \langle f_b \rangle \rangle_r$ and the standard deviation $\langle \sigma_b \rangle_r$ (for the distribution of MBRs $\{ f_b \}$) decrease (increase), as shown in Figs.~\ref{fig:SBS2}(g1) and \ref{fig:SBS2}(g2), which may result in better (worse) individual dynamics.
Figure \ref{fig:SBS2}(h) shows a plot of the bursting measure $\langle M_b \rangle_r$ versus $\Delta l$. With decreasing $\Delta l$ from 0, $\langle M_b \rangle_r$ decreases because both individual dynamics and efficiency of communication between nodes are worse. On the other hand, as $\Delta l$ is increased from 0, $\langle M_b \rangle_r$ increases mainly because of better individual dynamics overcoming worse efficiency of communication.

\subsection{Effects of Additive STDP on SBS}
\label{subsec:ASTDP}
We study the effect of additive STDP on SBS. The initial values of synaptic strengths $\{ J_{ij} \}$ are chosen from the Gaussian distribution where the mean $J_0$ is 2.5 and the standard
deviation $\sigma_0$ is 0.02. Then, $J_{ij}$ for each synapse is updated according to the additive nearest-burst pair-based STDP rule of Eq.~(\ref{eq:ASTDP}), in
contrast to the static case without STDP in Subsec.~\ref{subsec:NSTDP}.

\begin{figure}
\includegraphics[width=0.7\columnwidth]{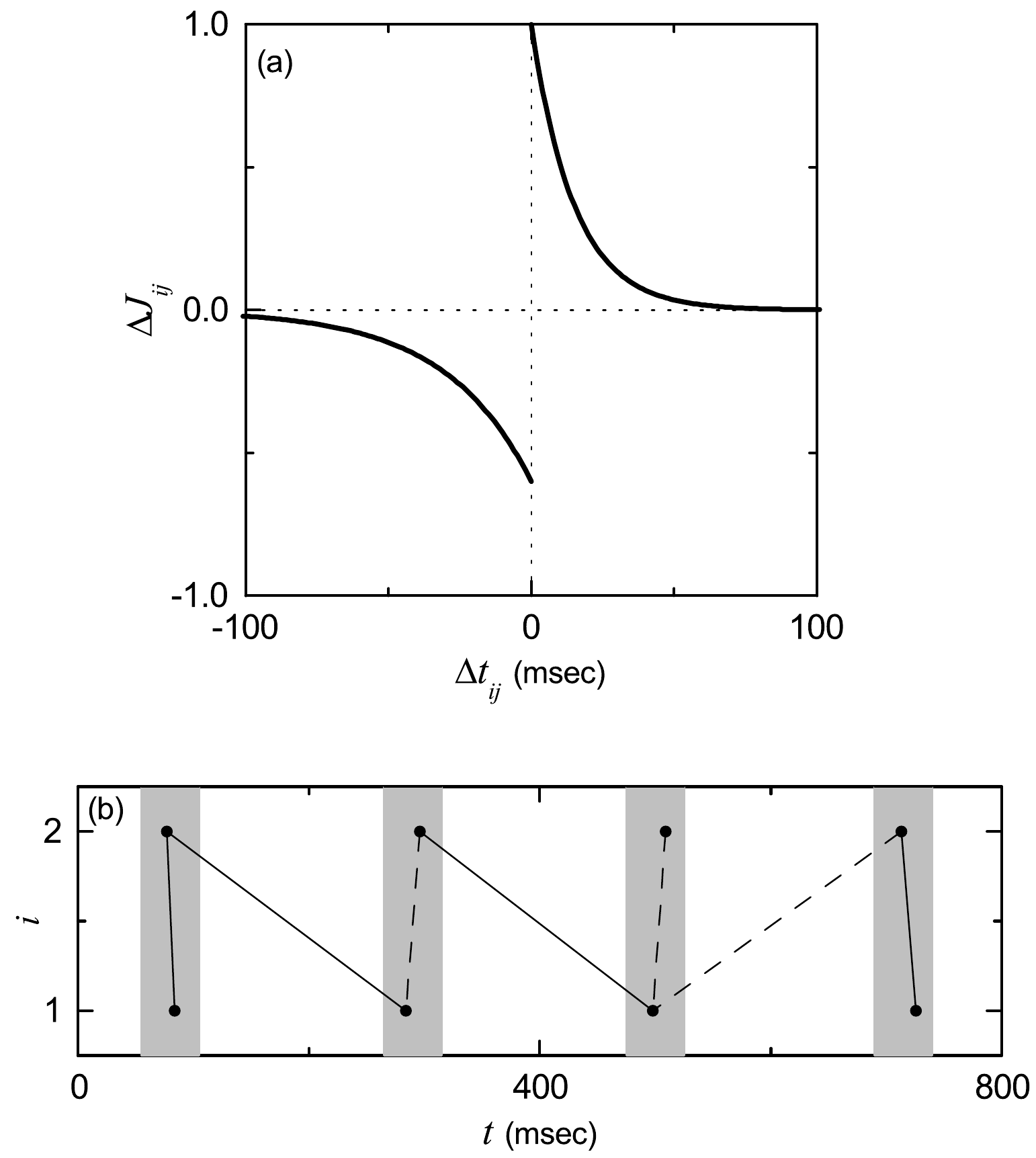}
\caption{(a) Time window for the Hebbian STDP. Plot of synaptic modification $\Delta J_{ij}$ versus $\Delta t_{ij}$ $(=t_i^{(post)} - t_j^{(pre)})$ for $A_+=1$, $A_{-}=0.6$, $\tau_+=15$ msec and $\tau_{-}=30$ msec.
 (b) Schematic diagram for the nearest-burst pair-based STDP rule; $i=1$ and 2 correspond to the post- and the pre-synaptic bursting neurons. Gray boxes and solid circles denote bursting stripes and burst onset times, respectively. Solid and dashed lines denote LTP and LTD, respectively.
}
\label{fig:TW}
\end{figure}

Figure \ref{fig:TW}(a) shows the time window for the synaptic modification $\Delta J_{ij}$ of Eq.~(\ref{eq:TW}) (i.e., plot of $\Delta J_{ij}$ versus $\Delta t_{ij}$). Here, $\Delta J_{ij}$ varies depending on the
relative time difference $\Delta t_{ij}$ $(=t_i^{(post)} - t_j^{(pre)})$ between the nearest burst onset times of the post-synaptic neuron $i$ and the pre-synaptic neuron $j$ \cite{Tass1,SS}. When a post-synaptic burst onset time
follows a pre-synaptic burst onset time (i.e., $\Delta t_{ij}$ is positive), LTP of synaptic strength appears; otherwise (i.e., $\Delta t_{ij}$ is negative), LTD occurs. A schematic diagram for the
nearest-burst pair-based STDP rule is given in Fig.~\ref{fig:TW}(b), where $i=1$ and 2 correspond to the post- and the pre-synaptic neurons, respectively. Here, gray boxes represent bursting stripes in
the raster plot, and burst onset times in the bursting stripes are denoted by solid circles. When the post-synaptic neuron ($i=1$) fires a bursting, LTP (denoted by solid lines) occurs via STDP between the
post-synaptic burst onset time and the previous nearest pre-synaptic burst onset time. In contrast, when the pre-synaptic neuron ($i=2$) fires a bursting, LTD (represented by dashed lines) occurs through STDP
between the pre-synaptic burst onset time and the previous nearest post-synaptic bust onset time. We note that such LTP/LTD may occur between the pre- and the post-synaptic burst onset times in the same bursting stripe or
in the different nearest-neighboring bursting stripes; solid/dashed lines connect pre- and post-synaptic burst onset times in the same or in the different nearest-neighboring bursting stripes.

\begin{figure}
\includegraphics[width=\columnwidth]{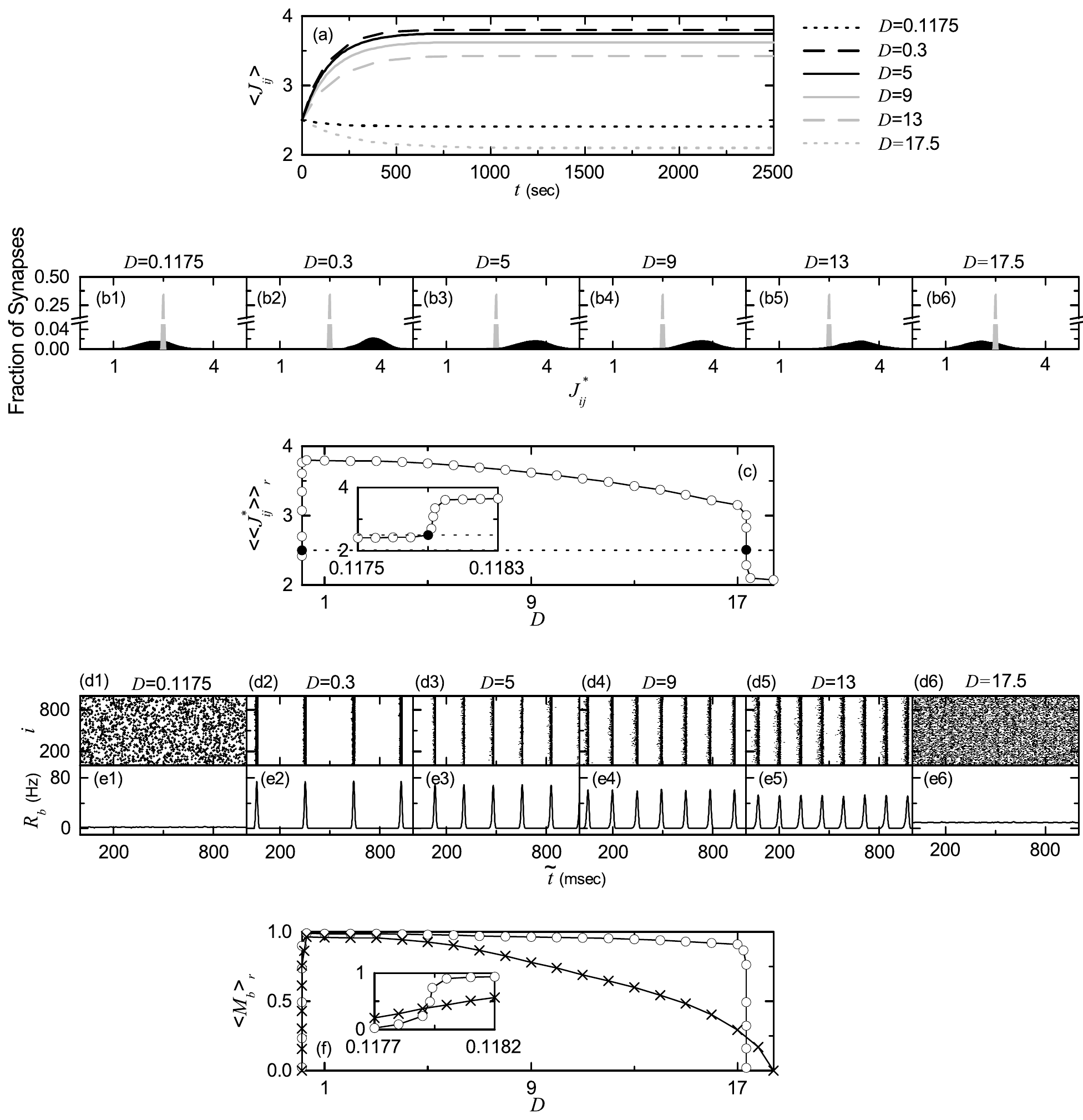}
\caption{Effect of additive STDP on SBS for the case of symmetric attachment with $l^*=10$; $N=10^3$. (a) Time-evolutions of population-averaged synaptic strengths $\langle J_{ij} \rangle$ for various values of $D$. (b1)-(b6) Histograms for the fraction of synapses versus $J^*_{ij}$ (saturated limit values of $J_{ij}$) are shown in black color for various values of $D$; for comparison, initial distributions of synaptic strengths
$\{ J_{ij} \}$ are also shown in gray color. (c) Plot of population-averaged limit values of synaptic strengths $\langle \langle J^*_{ij} \rangle \rangle_r$ versus $D$.
Raster plots of burst onset times in (d1)-(d6) and IPBR kernel estimates $R_b(t)$ in (e1)-(e6) for various values of $D$ after the saturation time, where $t=t^*$ (saturation time) + $\widetilde{t}$. (f) Plot of the statistical-mechanical bursting measure $\langle M_b \rangle_r$ (represented by open circles) versus $D$ in the saturated limit case. For comparison, $\langle M_b \rangle_r$ in the absence of STDP are also shown in crosses.
}
\label{fig:STDP1}
\end{figure}

Figure \ref{fig:STDP1}(a) shows time-evolutions of population-averaged synaptic strengths $\langle J_{ij} \rangle$ for various values of $D$ for the case of symmetric attachment with $l^*=10$;
$\langle \cdots \rangle$ represents an average over all synapses. In each case of $D=0.3,$ 5, 9 and 13, $\langle J_{ij} \rangle$ increases monotonically above its initial value $J_0$ (=2.5), and it approaches a saturated limit value $\langle J_{ij}^* \rangle$ nearly at $t=2000$ sec. As a result, LTP occurs for these values of $D$. On the other hand, for $D=0.1175$ and 17.5 $\langle J_{ij} \rangle$ decreases monotonically below $J_0$, and
converges to a saturated limit value $\langle J_{ij}^* \rangle$. Consequently, LTD takes place for these values of $D$.
Figures~\ref{fig:STDP1}(b1)-\ref{fig:STDP1}(b6) show histograms for fraction of synapses versus $J_{ij}^*$ (saturated limit values of $J_{ij}$ at $t=2000$ sec) in black color for various values of $D$;
the bin size for each histogram is 0.02. For comparison, initial distributions of synaptic strengths $\{ J_{ij} \}$ (i.e., Gaussian distributions whose mean $J_0$ and standard deviation $\sigma_0$ are 2.5 and 0.02, respectively) are also shown in gray color. For the cases of LTP ($D=0.3,$ 5, 9 and 13), their black histograms are located on the right side of the initial gray histograms, and hence their population-averaged values $\langle J_{ij}^*
\rangle$ become larger than the initial value $J_0$ (=2.5). On the other hand, the black histograms for the cases of LTD ($D=0.1175$ and 17.5) are shifted to the left side of the initial gray histograms, and hence their population-averaged values $\langle J_{ij}^* \rangle$ become smaller than $J_0$. For both cases of LTP and LTD, their black histograms are so much wider than the initial gray histogram [i.e., the standard deviations
$\sigma$ are very larger than the initial one $\sigma_0$ (=0.02)]; for clear views of broad black histograms, ``breaks'' are inserted on the vertical axes.
Figure \ref{fig:STDP1}(c) shows a plot of population-averaged limit values of synaptic strengths $\langle \langle J_{ij}^* \rangle \rangle_r$ versus $D$. Here, the horizontal dotted line denotes the initial average
value of coupling strengths $J_0$ (= 2.5), and the lower and the higher threshold values $\widetilde{D}_l$ $(\simeq 0.1179)$ and $\widetilde{D}_h$ $(\simeq 17.336)$ for LTP/LTD
(where $\langle \langle J_{ij}^* \rangle \rangle_r = J_0$) are represented by solid circles. Hence, LTP occurs in the range of ($\widetilde{D}_l$, $\widetilde{D}_h$); otherwise, LTD appears. We note that the range of ($\widetilde{D}_l$, $\widetilde{D}_h$) is strictly contained in the range of ($D^*_l$, $D^*_h$) ($D^*_l \simeq 0.1173$ and $D^*_h \simeq 18.4$) where SBS appears in the absence of STDP. Therefore, in most range of the SBS,
LTP occurs, while LTD takes place only near both ends.

We now consider the effects of LTP/LTD on SBS after the saturation time $t^*$ (= 2000 sec) in the case of symmetric attachment with $l^*=10$. Burst synchronization may be well visualized in the raster plot of bust onset times, and the corresponding IPBR kernel estimate $R_b(t)$ shows the population bursting behaviors well. Figures \ref{fig:STDP1}(d1)-\ref{fig:STDP1}(d6) and Figures \ref{fig:STDP1}(e1)-\ref{fig:STDP1}(e6) show raster plots of burst onset times and the corresponding IPBR kernel estimates $R_b(t)$ for various values of $D$, respectively.
In comparison with Figs.~\ref{fig:SBS1}(e1)-\ref{fig:SBS1}(e6) and Figs.~\ref{fig:SBS1}(f1)-\ref{fig:SBS1}(f6) in the absence of STDP, the degree of SBS for the case
of LTP ($D=0.3$ 5, 9 and 13) is increased so much. On the other hand, for the case of LTD ($D=0.1175$ and 17.5) the population states become desynchronized.
We also characterize the SBS in terms of the statistical-mechanical bursting measure $M_b$ of Eq.~(\ref{eq:BM}). Figure \ref{fig:STDP1}(f) shows the plot of $\langle M_b \rangle_r$ (denoted by open circles) versus $D$; for  comparison, $\langle M_b \rangle_r$ in the absence of STDP is also shown in crosses. A Matthew effect in synaptic plasticity occurs via a positive feedback process. Good burst synchronization with higher $M_b$ gets better via LTP, while bad burst synchronization with lower $M_b$ gets worse via LTD. As a result, a rapid step-like transition to SBS occurs, in contrast to the relatively smooth transition in the absence of STDP.

\begin{figure}
\includegraphics[width=\columnwidth]{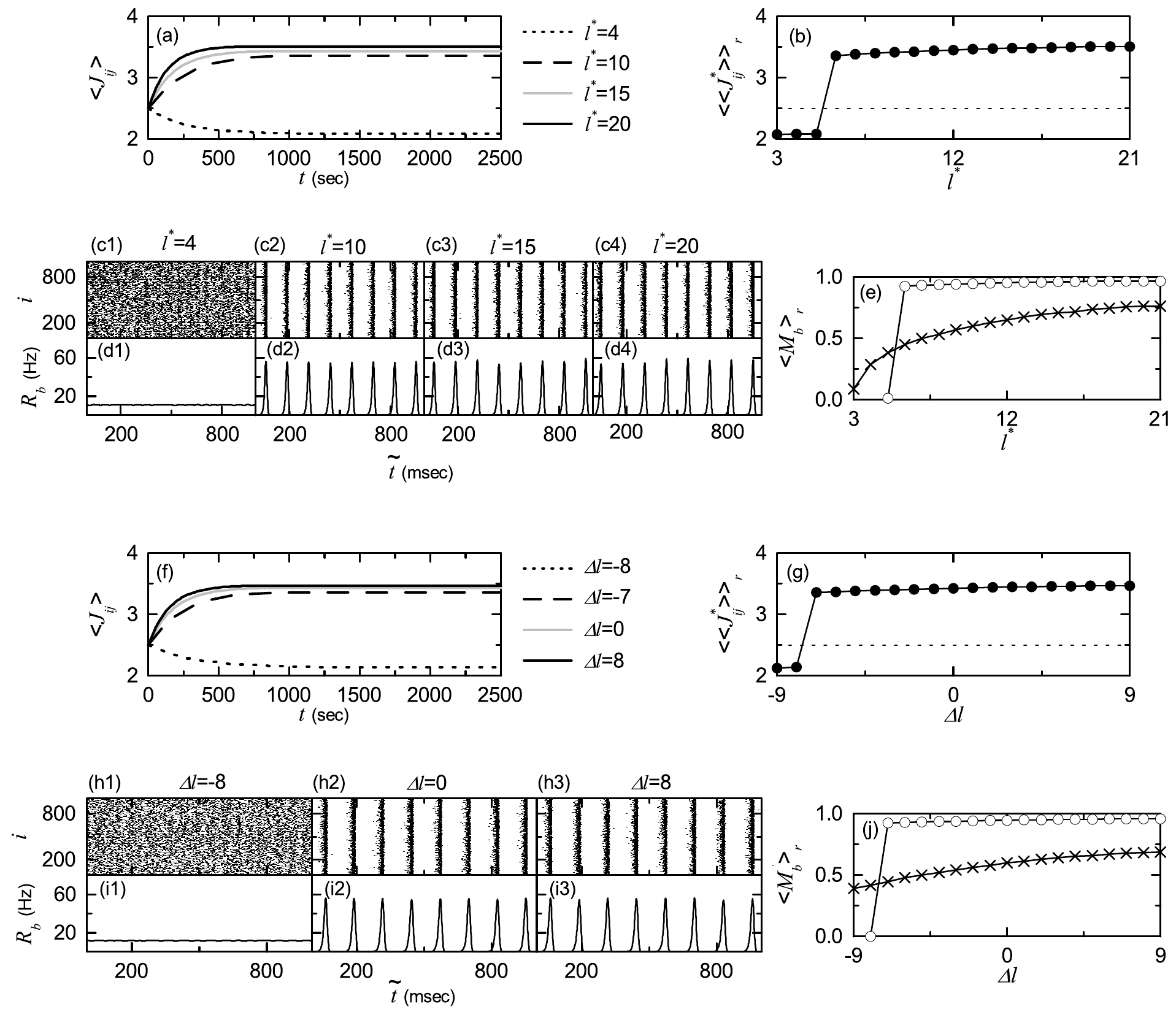}
\caption{Effect of network architecture on SBS in the presence of additive STDP for $D=13$; $N=10^3$.
Symmetric preferential attachment with $l_{in} = l_{out} = l^*$. (a) Time-evolutions of population-averaged synaptic strengths $\langle J_{ij} \rangle$ for various values of $l^*$. (b) Plot of population-averaged limit values of synaptic strengths $\langle \langle J^*_{ij} \rangle \rangle_r$ ($J^*_{ij}:$ saturated limit values of $J_{ij}$) versus $l^*$.
Raster plots of burst onset times in (c1)-(c4) and IPBR kernel estimates $R_b(t)$ in (d1)-(d4) for various values of $l^*$ after the saturation time, where $t=t^*$ (saturation time) + $\widetilde{t}$. (e) Plot of the statistical-mechanical bursting measure $\langle M_b \rangle_r$ (represented by open circles) versus $l^*$ in the saturated limit case. Asymmetric preferential attachment with $l_{in}= l^* + \Delta l$ and $l_{out} = l^* - \Delta l$ ($l^*=10$) (f) Time-evolutions of population-averaged synaptic strengths   $\langle J_{ij} \rangle$  for various values of $\Delta l$. (g) Plot of population-averaged limit values of synaptic strengths $\langle \langle J^*_{ij} \rangle \rangle_r$ versus $\Delta l$. Raster plots of burst onset times in (h1)-(h3) and IPBR kernel estimates $R_b(t)$ in (i1)-(i3) for various values of $\Delta l$ after the saturation time, where $t=t^*$ (saturation time) + $\widetilde{t}$. (j) Plot of the statistical-mechanical bursting measure $\langle M_b \rangle_r$ (represented by open circles) versus $\Delta l$ in the saturated limit case. For comparison, $\langle M_b \rangle_r$ in the absence of STDP are also shown in crosses in (e) and (j).
}
\label{fig:STDP2}
\end{figure}

The effect of scale-free connectivity on SBS for the static case of fixed coupling strengths is studied for $D=13$ by varying the degree of symmetric attachment $l^*$ and the asymmetry parameter
$\Delta l$, and the results in the absence of STDP are shown in Fig.~\ref{fig:SBS2}. From now on, we take into consideration the synaptic plasticity and investigate the effect of network architecture
on the SBS for $D=13$ in both cases of symmetric and asymmetric attachments by changing $l^*$ and $\Delta l$, respectively.
We first consider the case of symmetric attachment (i.e., $l_{in}=l_{out}  = l^*$).
Figure \ref{fig:STDP2}(a) shows time-evolutions of population-averaged synaptic strengths $\langle J_{ij} \rangle$ for various values of $l^*$.
For each case of $l^*=6,$ 10, and 20, $\langle J_{ij} \rangle$ increases monotonically above its initial value $J_0$ (=2.5), and it converges toward a saturated limit value $\langle
J_{ij}^* \rangle$ nearly at $t=2000$ sec. Consequently, LTP occurs for these values of $l^*$. In contrast, for $l^*=4$ $\langle J_{ij} \rangle$ decreases monotonically below $J_0$, and
approaches a saturated limit value $\langle J_{ij}^* \rangle$. Accordingly, for this case LTD takes place. Figure \ref{fig:STDP2}(b) shows a plot of population-averaged limit values of synaptic strengths $\langle \langle J_{ij}^* \rangle \rangle_r$ versus $l^*$; the horizontal dotted line represents the initial average value of coupling strengths $J_0$ (= 2.5). For $l^* \geq 6$ LTP occurs, while for $l^* \leq 5$ LTD takes place.
We also consider the effects of LTP/LTD on the SBS after the saturation time $t^*$ (= 2000 sec). Figures \ref{fig:STDP2}(c1)-\ref{fig:STDP2}(c4) and Figures \ref{fig:STDP2}(d1)-\ref{fig:STDP2}(d4) show raster plots of burst onset times and the corresponding IPBR kernel estimates $R_b(t)$ for various values of $l^*$, respectively. The degrees of SBS for the case of LTP ($l^* =10,$ 15, and 20) are increased so much when compared with Figs.~\ref{fig:SBS2}(a3)-\ref{fig:SBS2}(a5) and Figs.~\ref{fig:SBS2}(b3)-\ref{fig:SBS2}(b5) in the absence of STDP.  In contrast, for the case of LTD ($l^*=4$) the population states become desynchronized.
The SBS is characterized in terms of the statistical-mechanical bursting measure $M_b$ of Eq.~(\ref{eq:BM}). Figure \ref{fig:STDP2}(e) shows the plot of $\langle M_b \rangle_r$ (denoted by open circles) versus $l^*$; for  comparison, $\langle M_b \rangle_r$ in the absence of STDP is also shown in crosses. Like the case in Fig.~\ref{fig:STDP1}(f), a Matthew effect in synaptic plasticity occurs via a positive feedback process. Thus, good burst synchronization with higher $M_b$ gets better via LTP, while bad burst synchronization with lower $M_b$ gets worse via LTD. Consequently, a rapid step-like transition to SBS occurs, in contrast to the relatively smooth transition in the absence of STDP.

\begin{figure}
\includegraphics[width=\columnwidth]{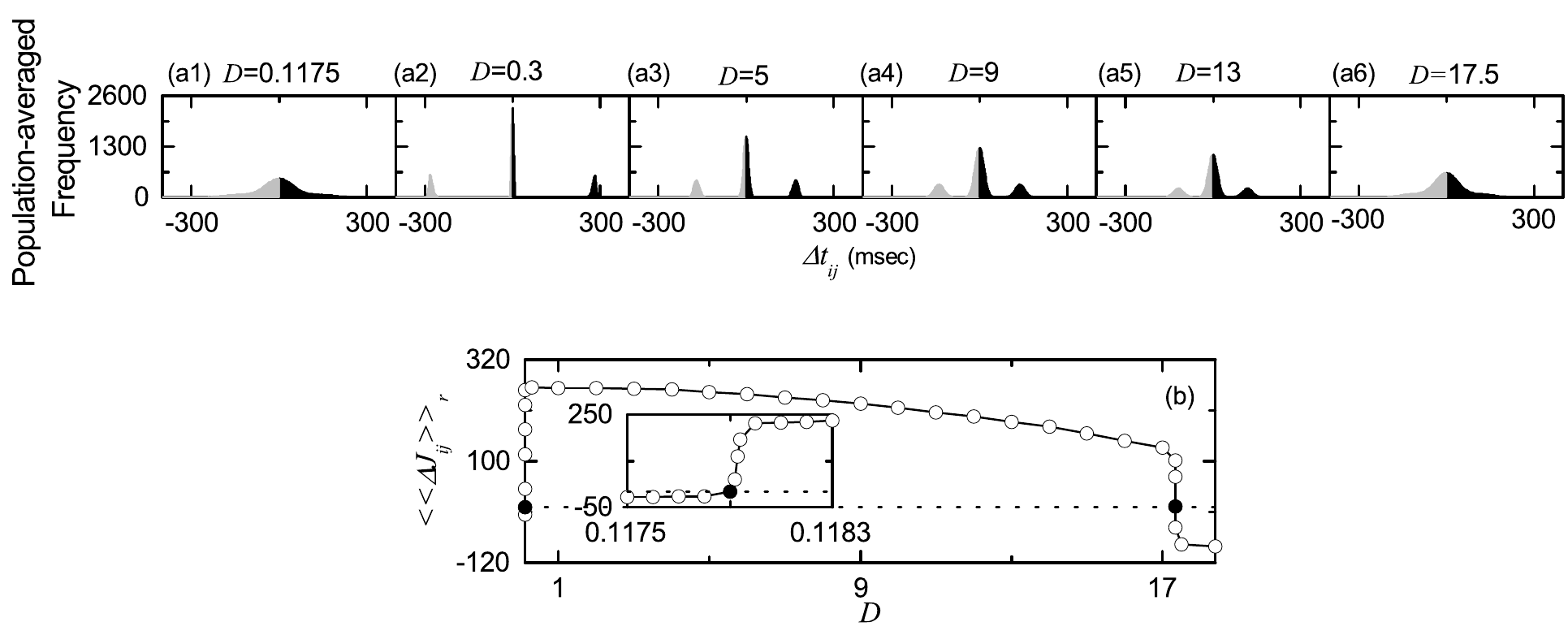}
\caption{Distributions of microscopic time delays $\{ \Delta t_{ij} \}$ between the pre- and the post-synaptic burst onset times and synaptic modifications for the case of symmetric attachment with $l^*=10$; $N=10^3$. (a1)-(a6) Population-averaged histograms $H(\Delta t_{ij})$ for the distributions of time delays $\{ \Delta t_{ij} \}$ during the time interval from $t=0$ to the saturation time $t^*$ (=2000 sec) for various values of $D$; black and gray regions represent LTP and LTD, respectively. (b) Plot of the population-averaged synaptic modifications $\langle \langle \Delta J_{ij} \rangle \rangle_r$ [during the time interval from $t=0$ to the saturation time $t^*$ (=2000 sec)] versus $D$. The values of $\langle \langle \Delta J_{ij} \rangle \rangle_r$ are obtained from the population-averaged histograms $H(\Delta t_{ij})$ in (a).
}
\label{fig:MI1}
\end{figure}

Next, we consider the case of asymmetric attachment [i.e., $l_{in}= l^* + \Delta l$ and $l_{out}= l^* - \Delta l$ ($l^*=10$)].
Time-evolutions of population-averaged synaptic strengths $\langle J_{ij} \rangle$ for various values of $\Delta l$ are shown
in Fig.~\ref{fig:STDP2}(f). In each case of $\Delta l=-7,$ 0, and 8, $\langle J_{ij} \rangle$ increases monotonically above its initial value $J_0$ (=2.5), and it approaches a saturated limit value $\langle
J_{ij}^* \rangle$ nearly at $t=2000$ sec. As a result, LTP occurs for these values of $l^*$. On the other hand, for $\Delta l=-8,$ $\langle J_{ij} \rangle$ decreases monotonically below $J_0$, and
converges toward a saturated limit value $\langle J_{ij}^* \rangle$. Accordingly, for this case LTD takes place.
A plot of population-averaged limit values of synaptic strengths $\langle \langle J_{ij}^* \rangle \rangle_r$ versus $\Delta l$ is shown in Figure \ref{fig:STDP2}(g);
the horizontal dotted line represents the initial average value of coupling strengths $J_0$ (= 2.5).
For $\Delta l \geq -7$ LTP occurs, while for $\Delta l \leq -8$ LTD takes place. We consider the effects of LTP/LTD on the SBS after the saturation time $t^*$ (= 2000 sec). Figures \ref{fig:STDP2}(h1)-\ref{fig:STDP2}(h3) and Figures \ref{fig:STDP2}(i1)-\ref{fig:STDP2}(i3) show raster plots of burst onset times and the corresponding IPBR kernel estimates $R_b(t)$ for various values of $\Delta l$, respectively. The degrees of SBS for the case of LTP ($\Delta l =0$ and 8) are increased so much when compared with Figs.~\ref{fig:SBS2}(e2)-\ref{fig:SBS2}(e3) and Figs.~\ref{fig:SBS2}(f2)-\ref{fig:SBS2}(f3) in the absence of STDP.  On the other hand, in the case of LTD
($\Delta l=-8$) the population state becomes desynchronized. We also characterize the SBS in terms of the statistical-mechanical bursting measure $M_b$. Figure \ref{fig:STDP2}(j) shows the plot of $\langle M_b \rangle_r$ (denoted by open circles) versus $\Delta l$; for  comparison, $\langle M_b \rangle_r$ in the absence of STDP is also shown in crosses. As in the case in Fig.~\ref{fig:STDP2}(e), a Matthew effect in synaptic plasticity occurs via a positive feedback process. Hence, good burst synchronization with higher $M_b$ gets better via LTP, while bad burst synchronization with lower $M_b$ gets worse via LTD. As a result, a rapid step-like transition to SBS occurs, in contrast to the relatively smooth transition in the absence of STDP.

From now on, we consider the case of symmetric attachment with $l^*=10$, and investigate emergences of LTP and LTD of synaptic strengths intensively through our own microscopic methods based on the distributions of time delays $\{ \Delta t_{ij} \}$ between the pre- and the post-synaptic burst onset times. Population-averaged histograms $H(\Delta t_{ij})$ for the distributions of time delays $\{ \Delta t_{ij} \}$ are shown in Figs.~\ref{fig:MI1}(a1)-\ref{fig:MI1}(a6) for various values of $D$: for each synaptic pair, its histogram for the distribution of $\{ \Delta t_{ij} \}$ during the time interval from $t=0$ to the saturation time
$t^*$ (=2000 sec) is obtained, and then we get the population-averaged histogram through averaging over all synaptic pairs. Black and gray regions in the histograms denote LTP and LTD, respectively.
For the case of LTP ($D=0.3,$ 5, 9, and 13), there exist 3 peaks in each histogram: one main central peak and two left and right minor peaks.
When the pre- and the post-synaptic burst onset times appear in the same bursting stripe in the raster plot of burst onset times, its time delay $\Delta t_{ij}$ lies in the main peak.
For this case, LTP/LTD may occur depending on the sign of $\Delta t_{ij}$; for $\Delta t_{ij}>0~ (<0)$, LTP (LTD) takes place.
In contrast, time delay $\Delta t_{ij}$ lies in the minor peak when the pre- and the post-synaptic burst onset times appear in the different nearest-neighboring bursting stripes.
If the pre-synaptic (post-synaptic) bursting stripe precedes the post-synaptic (pre-synaptic) bursting stripe, then its time delay $\Delta t_{ij}$ lies in the right (left) minor peak; LTP (LTD) occurs
in the right (left) minor peak. However, for the case of LTD ($D=0.1175$ and 17.5), the population states become desynchronized due to overlap of bursting stripes in the raster plot of burst onset times.
As a result, the main peak in the histogram becomes merged with the left and the right minor peaks, and then only one broadened single peak appears, in contrast to the case of LTP
($D=0.3,$ 5, 9, and 13). Then, the population-averaged synaptic modification $\langle \langle \Delta J_{ij} \rangle \rangle_r$ [during the time interval from $t=0$ to the saturation time $t^*$ (=2000 sec)]
may be directly obtained from the above histogram $H(\Delta t_{ij})$:
\begin{equation}
  \langle \langle \Delta J_{ij} \rangle \rangle_r \simeq \sum_{\rm {bins}} H(\Delta t_{ij}) \cdot \Delta J_{ij} (\Delta t_{ij}).
\end{equation}
A plot of $\langle \langle \Delta J_{ij} \rangle \rangle_r$ is shown in Fig.~\ref{fig:MI1}(b). Here, solid circles represent the lower and the higher thresholds $\widetilde{D}_l$ and $\widetilde{D}_h$ for
LTP/LTD (where $\langle \langle \Delta J_{ij} \rangle \rangle_r = 0$), which are the same as those in Fig.~\ref{fig:STDP1}(c). LTP occurs in the range of ($\widetilde{D}_l$, $\widetilde{D}_h$) because
$\langle \langle \Delta J_{ij} \rangle \rangle_r >0$, while LTD appears in the remaining region where $\langle \langle \Delta J_{ij} \rangle \rangle_r < 0$. Then, population-averaged saturated limit values of
synaptic strengths $\langle \langle J_{ij}^* \rangle \rangle_r$ (given by $J_0 + \delta~\langle \langle \Delta J_{ij} \rangle \rangle_r$) agree well with the directly-obtained values in Fig.~\ref{fig:STDP1}(c).

\begin{figure}
\includegraphics[width=\columnwidth]{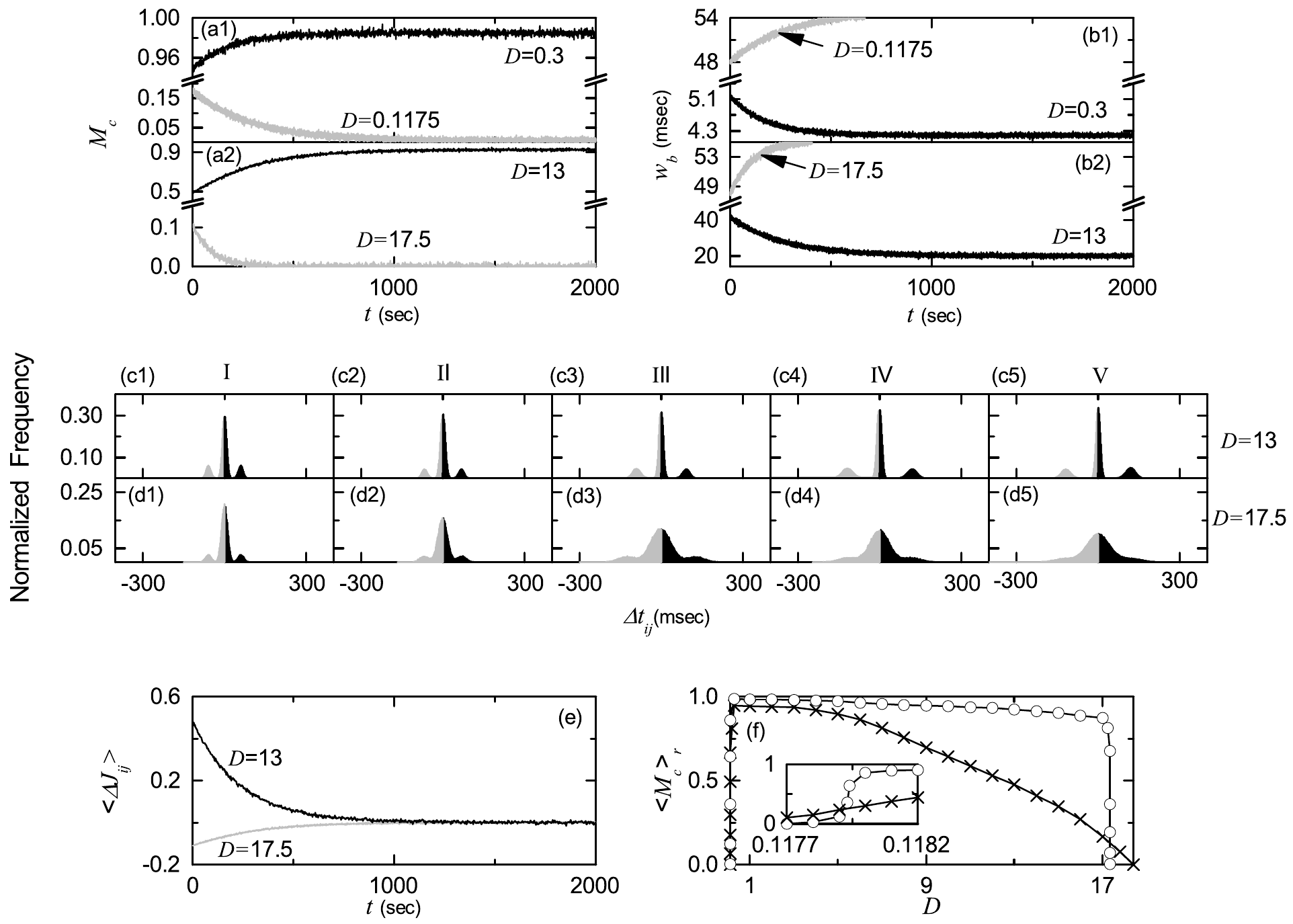}
\caption{Microscopic pair-correlations for the case of symmetric attachment with $l^*=10$; $N=10^3$. Time-evolutions of the microscopic correlation measures $M_c(t)$ for (a1) $D =$ 0.1175 and 0.3 and (a2) $D =$ 13 and 17.5.
Time-evolutions of the widths $w_b(t)$ of the bursting stripes in the raster plot of burst onset times for (b1) $D =$ 0.1175 and 0.3 and (b2) $D =$ 13 and 17.5.
Time-evolutions of the normalized histograms $H(\Delta t_{ij})$ for the distributions of time delays $\{ \Delta t_{ij} \}$ between the pre- and the post-synaptic birst onset times for $D = 13$ in (c1)-(c5) and for $D = 17.5$ in (d1)-(d5); 5 stages are shown in I (starting from $\sim$ 0 sec), II (starting from $\sim$ 100 sec), III (starting from $\sim$ 300 sec), IV (starting from $\sim$ 500 sec), and V (starting from $\sim$ 1000 sec). (e) Time-evolutions of population-averaged synaptic modifications $\langle \Delta J_{ij}(t) \rangle$ for $D = 13$ (black line) and for $D = 17.5$ (gray line). (f) Plot of $\langle M_c \rangle_r$ (represented by open circles) versus $D$ in the saturated limit case. For comparison, $\langle M_c \rangle_r$ in the absence of STDP are also shown in crosses.
}
\label{fig:MI2}
\end{figure}

Finally, in the case of symmetric attachment with $l^*=10$, we investigate the effect of STDP on the microscopic dynamical pair-correlation $C_{ij}(\tau)$ between the pre- and the post-synaptic IIBRs (instantaneous individual
burst rates) for the $(i,j)$ synaptic pair. Each train of burst onset times for the $i$th neuron is convoluted with a Gaussian kernel function $K_h(t)$ of band width $h$ to get a smooth estimate of IIBR $r_i(t)$:
\begin{equation}
r_i(t) = \sum_{b=1}^{n_i} K_h (t-t_{b}^{(i)}),
\label{eq:IISR}
\end{equation}
where $t_{b}^{(i)}$ is the $b$th burst onset time of the $i$th neuron, $n_i$ is the total number of burst onset times for the $i$th neuron, and $K_h(t)$ is given in Eq.~(\ref{eq:Gaussian}).
Then, the normalized temporal cross-correlation function $C_{ij}(\tau)$ between the IIBR kernel estimates $r_i(t)$ and $r_j(t)$ of the $(i,j)$ synaptic pair is given by:
\begin{equation}
C_{ij}(\tau) = \frac{\overline{\Delta r_i(t+\tau) \Delta r_j(t)}}{\sqrt{\overline{\Delta r^2_i(t)}}\sqrt{\overline{\Delta r^2_j(t)}}},
\end{equation}
where $\Delta r_i(t) = r_i(t) - \overline{r_i(t)}$ and the overline denotes the time average.
Then, the microscopic correlation measure $M_c,$ representing the average ``in-phase'' degree between the pre- and the post-synaptic pairs, is given by the average value of $C_{ij}(0)$ at the zero-time
lag for all synaptic pairs:
\begin{equation}
M_c = \frac{1}{N_{syn}} \sum_{(i,j)} C_{ij}(0),
\label{eq:CM}
\end{equation}
where $N_{syn}$ is the total number of synapses.
Time-evolutions of the microscopic correlation measures $M_c(t)$ for the population states are shown in Figs.~\ref{fig:MI2}(a1)-\ref{fig:MI2}(a2). Data for calculation of $M_c(t)$ are obtained through
averages during successive 5 global cycles of the IPBR kernel estimate $R_b(t)$ for both cases of LTP and LTD.
In Fig.~\ref{fig:MI2}(a1), we consider two small values of $D$ (= 0.3 and 0.1175 corresponding to the cases of LTP and LTD, respectively).
The initial values of $M_c$ for $D=0.3$ and 0.1175 are 0.95 and 0.17, respectively. With increase in time $t$, $M_c$ for $D=0.3$ increases, and it approaches a limit value ($M_c=0.99$). In contrast, $M_c$ for $D=0.1175$
decreases with time $t$, and it seems to converge toward zero.
Similarly, we also consider two large values of $D$ (= 13 and 17.5 corresponding to the cases of LTP and LTD, respectively) in Fig.~\ref{fig:MI2}(a2).
As the time $t$ increases, $M_c$ for $D=13$ increases to a limit value ($M_c=0.92$), while $M_c$ for $D=17.5$ tends to decrease to zero.
Enhancement (suppression) in $M_c$ leads to increase (decrease) in the average in-phase degree between the pre- and the post-synaptic pairs. Then, widths of bursting stripes in the raster plot of
burst onset times decrease (increase) due to enhancement (suppression) of $M_c$. Time-evolutions of the width $w_b(t)$ of the bursting stripes are shown in Fig.~\ref{fig:MI2}(b1)-\ref{fig:MI2}(b2).
Here, $w_b(t)$ is obtained through averaging the widths of bursting stripes during successive 5 global cycles of $R_b(t)$. For $D=0.3$ and 13, $w_b(t)$ decreases due to enhancement in $M_c$, which results in narrowed distribution of time delays $\{ \Delta t_{ij} \}$ between the pre- and the post-synaptic burst onset times. As a result, LTP may occur. On the other hand, for $D=0.1175$ and 17.5, $w_b(t)$ increases due to suppression in $M_c$
(calculations of $w_b(t)$ for $D=$ 0.1175 and 17.5 are made until $t \simeq$ 669 sec and 406 sec, respectively, when bursting stripes begin to overlap), which leads to widened distribution of time delays $\{ \Delta t_{ij} \}$.
Consequently, LTD may take place.

Figures \ref{fig:MI2}(c1)-\ref{fig:MI2}(c5) for $D= 13$ and Figs.~\ref{fig:MI2}(d1)-\ref{fig:MI2}(d5) for $D=17.5$ show time-evolutions of normalized histograms $H(\Delta t_{ij})$ for the distributions of time delays $\{ \Delta t_{ij} \}$; the bin size in each histogram is 2 msec.
Here, we consider 5 stages, represented by I ($12 \sim 342$ msec for $D=13$ and $15 \sim 315$ msec for $D=17.5$), II ($100008 \sim  100428$ msec for $D=13$  and $100012 \sim  100302$ msec for $D=17.5$), III ($300012 \sim  300532$ msec for $D=13$, and $300002 \sim  300287$ msec for $D=17.5$), IV ($500004 \sim  500624$ msec for $D=13$ and $500005 \sim  500285$ msec for $D=17.5$), and  V ($1000006 \sim  1000646$ msec for $D=13$ and $1000002 \sim  1000282$ msec for $D=17.5$).
At each stage, we obtain the distribution for $\{ \Delta t_{ij} \}$ for all synaptic pairs during the 5 global cycles of the IPBR $R_b(t)$ and get the normalized histogram by dividing the
distribution with the total number of synapses (=20000). For the case of $D=13$ (LTP), 3 peaks appear in each histogram; main central peak and two left and right minor peaks. With increase in time $t$ (i.e., with increasing the level of stage), peaks become narrowed, and then they become sharper. The intervals between the main peak and the two minor peaks also increase a little because the bursting frequency $f_b$ of $R_b(t)$ decreases with the stage. Moreover, with increasing the stage, the main peak becomes more and more symmetric, and hence the effect of LTP in the black part tends to cancel out nearly the effect of LTD in the gray part at the stage V.
In the case of $D=17.5$ (LTD), as the level of the stage is increased, peaks become wider and the merging-tendency between the peaks is intensified. For the stages IV and V, only one broad central
peak seems to appear. At the stage V, the effect of LTP in the black part tends to nearly cancel out the effect of LTD in the gray part because the broad peak is nearly symmetric.
From these normalized histograms $H(\Delta t_{ij})$, we also obtain the population-averaged synaptic modification $\langle \Delta J_{ij} \rangle$ [$\simeq \sum_{\rm{bins}} H(\Delta t_{ij})
\cdot \Delta J_{ij} (\Delta t_{ij})$]. Figure \ref{fig:MI2}(e) shows time-evolutions of $\langle \Delta J_{ij} \rangle$ for $D=13$ (black curve) and $D=17.5$ (gray curve).
$\langle \Delta J_{ij} \rangle$ for $D=13$ is positive. On the other hand, it is negative for $D=17.5$.
For both cases, they converge toward nearly zero at the stage V $(t \sim 1000$ sec) because the normalized histograms become nearly symmetric.
Then, the time evolution of population-averaged synaptic strength $\langle J_{ij} \rangle$ is given by $\langle J_{ij} \rangle = J_0 + \delta \sum_k \langle \Delta J_{ij}(k)
\rangle,$ where $J_0$(initial average synaptic strength)= 2.5 and $k$ represents the average for the $k$th 5 global cycles of $R_b(t)$.
Time-evolutions of $\langle J_{ij} \rangle$ (obtained in this way) for $D=13$ and 17.5 agree well with directly-obtained ones in Fig.~\ref{fig:STDP1}(a). Consequently, LTP (LTD) occurs for $D=13$ (17.5).

Figure \ref{fig:MI2}(f) shows plots of $\langle M_c \rangle_r$ versus $D$ in the presence (open circles) and the absence (crosses) of STDP.
The number of data used for the calculation of each temporal cross-correlation function $C_{ij}(\tau)$ (the values of $C_{ij}(0)$ at the zero time lag are used for calculation of $M_c$) is $2^{16}$ (=65536) after the saturation time $t^*$ (=2000 sec) in each realization. As in the case of $\langle M_b \rangle_{r}$ in Fig.~\ref{fig:STDP1}(f), a Matthew effect also occurs in $\langle M_c \rangle_{r}$: good pair-correlation with higher $M_c$ gets better, while bad pair-correlation with lower $M_c$ gets worse. Hence, a step-like transition occurs, in contrast to the case without STDP.

\subsection{Effects of Multiplicative STDP on SBS}
\label{sec:MSTDP}
Here, we consider the case of symmetric attachment with $l^*=10$ and investigate the effect of multiplicative STDP (depending on states) on SBS in comparison with the (above) additive case (independent of states).
The coupling strength for each synapse is updated with a multiplicative nearest-burst pair-based STDP rule \cite{Tass1,Multi}:
\begin{equation}
J_{ij} \rightarrow J_{ij} + \delta~(J^*-J_{ij})~|\Delta J_{ij}(\Delta t_{ij})|.
\label{eq:MSTDP}
\end{equation}
Here, $\delta$ $(=0.005)$ is the update rate, $\Delta J_{ij}$ is the synaptic modification depending on the relative time difference $\Delta t_{ij}$ $(=t_i^{(post)} - t_j^{(pre)})$
between the nearest burst onset times of the post-synaptic neuron $i$ and the pre-synaptic neuron $j$ [time window for $\Delta J_{ij}$ is given in Eq.~(\ref{eq:TW})], and
$J^*=$ $J_h~(J_l)$ for the LTP (LTD) [$J_h(=5.0)$ and $J_l(=0.0001)$ is the higher (lower) bound of $J_{ij}$ (i.e., $J_{ij} \in [J_l, J_h])$].
For the case of multiplicative STDP, the bounds for the synaptic strength $J_{ij}$ become soft, because a change in synaptic strengths scales linearly with the distance to the
higher and the lower bounds, in contrast to hard bounds for the case of additive STDP.

\begin{figure}
\includegraphics[width=\columnwidth]{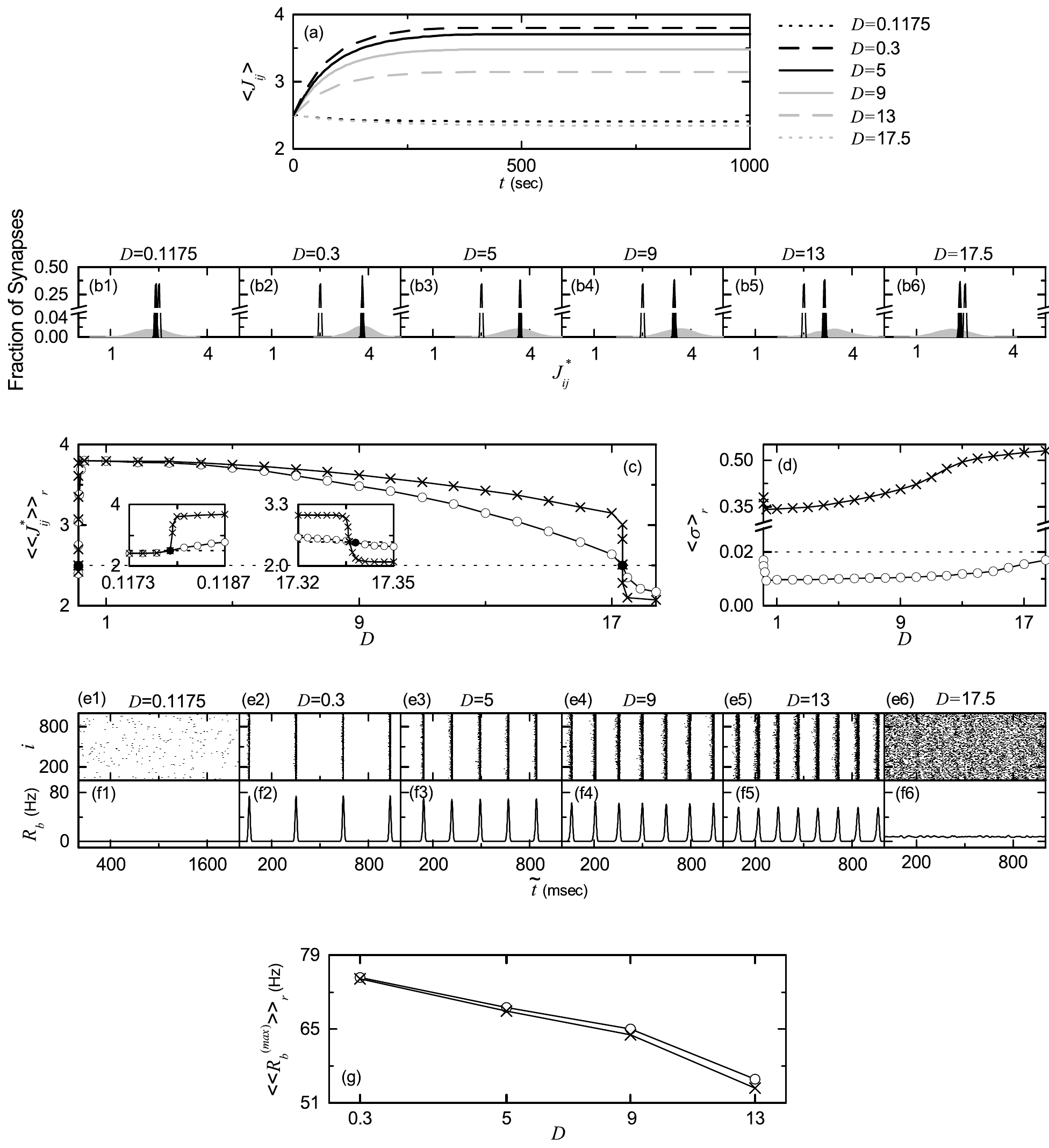}
\caption{Effects of multiplicative STDP on SBS for the case of symmetric attachment with $l^*=10$; $N=10^3$.
(a) Time-evolutions of population-averaged synaptic strengths $\langle J_{ij} \rangle$ for various values of $D$. (b1)-(b6) Histograms for the fraction of synapses versus
$J^*_{ij}$ (saturated limit values of $J_{ij}$) for various values of $D$ (black region); for comparison, distributions of $\{ J^*_{ij} \}$ for the case of additive STDP
and the initial distributions of $\{ J_{ij} \}$ are also shown in gray regions and in black curves, respectively.
(c) Plot of population-averaged limit values of synaptic strengths $\langle \langle J_{ij}^* \rangle \rangle_r$ (denoted by open circles) versus $D$.
For comparison, $\langle \langle J_{ij} \rangle \rangle_r$ in the case of additive STDP are also shown in crosses. (d) Plot of standard deviations $\langle \sigma \rangle_r$ (denoted by open circles) for the distribution of saturated limit coupling strengths $\{ J^*_{ij} \}$ versus $D$; for comparison, the values of $\langle \sigma \rangle_r$ in the case of additive STDP are also shown in crosses. Raster plots of burst onset times in (e1)-(e6) and IPBR kernel estimates $R_b(t)$ in (f1)-(f6) for various values of $D$
after the saturation time, where $t=t^*$ (saturation time) + $\widetilde{t}$. (g) Plot of average maximum values $\langle \langle R_b^{(max)} \rangle \rangle_r$ of the IPBR kernel estimate $R_b(t)$ (denoted by open circles) versus $D$ (=0.3, 5, 9, and 13); for comparison, values of $\langle \langle R_b^{(max)} \rangle \rangle_r$ for the case of additive STDP are represented in crosses.
}
\label{fig:MULTI1}
\end{figure}

Figure \ref{fig:MULTI1}(a) shows time-evolutions of population-averaged synaptic strengths $\langle J_{ij} \rangle$ for various values of $D$.
For $D=0.3,$ 5, 9, and 13, $\langle J_{ij} \rangle$ increases above its initial value $J_0$ (= 2.5), and converges toward a saturated limit value $\langle J_{ij}^* \rangle$ nearly at $t=500$ sec.
Consequently, LTP occurs for these values of $D$. In contrast, for $D=0.1175$ and 17.5 $\langle J_{ij} \rangle$ decreases below $J_0$, and approaches a saturated limit value $\langle J_{ij}^* \rangle$.
As a result, LTD occurs for these values of $D$. For this multiplicative case, the saturation time is shorter and deviations of the saturated limit values $J_{ij}^*$ from $J_0$ are generally (except for the case of small $D$) smaller due to the soft bounds, in comparison with the additive case in Fig.~\ref{fig:STDP1}(a); for small $D=0.1175$ and 0.3, the values of $J_{ij}^*$ are the same in both the additive and the multiplicative cases.

Histograms for fraction of synapses versus $J_{ij}^*$ (saturated limit values of $J_{ij}$ at $t=500$ sec) are shown in black regions for various values of $D$ in Figs.~\ref{fig:MULTI1}(b1)-\ref{fig:MULTI1}(b6);
the bin size for each histogram is 0.02. For comparison, distributions of $\{ J^*_{ij} \}$ for the case of additive STDP and initial Gaussian distributions (mean $J_0$ = 2.5 and standard deviation
$\sigma_0$ = 0.02) of $\{ J_{ij} \}$ are also shown in gray regions and in black curves, respectively; for clear views of gray wide histograms, breaks are inserted on the vertical axes.
Like the case of additive STDP, LTP occurs for $D=0.3,$ 5, 9, and 13, because their black histograms lie on the right side of the initial black-curve histograms.
The black histograms for the multiplicative case lie generally on the left side of the gray histograms for the case of additive STDP (except for the case of $D=0.3$ where
peaks of nearly symmetric distributions for both the additive and the multiplicative cases coincide nearly). Hence, the population-averaged values $\langle J_{ij}^* \rangle$ for the multiplicative case are generally
smaller than those for the additive case, due to soft bounds; in the exceptional case of $D=0.3$ $\langle J_{ij}^* \rangle$ is nearly the same for both the multiplicative and the additive cases.
Particularly, the black histograms for the multiplicative case (with soft bounds) are much narrower than the gray histograms for the additive case (with hard bounds). As a result, standard deviations $\sigma$ for distributions
of $\{ J^*_{ij} \}$ in the black histograms are much smaller than those for the additive case, because their variations in $J_{ij}$ are restricted due to soft bounds in comparison with hard bounds for the additive case.
These standard deviations $\sigma$ for the multiplicative case are even smaller than the initial ones $\sigma_0$ (= 0.02).
On the other hand, for $D=0.1175$ and 17.5 LTD occurs because the black histograms are shifted to the left side of the initial black-curve histograms.
But, the black histograms for the multiplicative case lie generally on the right side of the gray histograms for the case of additive STDP (except for the case of $D=0.1175$ where
peaks of nearly symmetric distributions for both the additive and the multiplicative cases become nearly the same). Hence, the population-averaged values $\langle J_{ij}^* \rangle$ for the multiplicative case are generally
larger than those for the additive case, due to soft bounds; for the exceptional case of $D=0.1175$ $\langle J_{ij}^* \rangle$ is nearly the same in both the multiplicative and the additive cases.
Like the case of LTP, the histograms for the multiplicative case are much narrower than those for the additive case. Consequently, standard deviations $\sigma$ for distributions of $\{ J^*_{ij} \}$ in the multiplicative case are much smaller than those for the additive case. Furthermore, these standard deviations $\sigma$ are even smaller than the initial ones $\sigma_0$ (=0.02), as in the case of LTP.

A plot of population-averaged limit values  $\langle \langle J_{ij}^* \rangle \rangle_r$ (denoted by open circles for the multiplicative case) of synaptic strengths versus $D$ is shown in Fig.~\ref{fig:MULTI1}(c).
Here, the horizontal dotted line represents the initial average value of coupling strengths $J_0$ (= 2.5), and the lower and the higher thresholds $\widetilde{D}_l^*$ $(\simeq 0.1179)$ and $\widetilde{D}_h^*$
$(\simeq 17.338)$ for LTP/LTD (where $\langle \langle J_{ij}^* \rangle \rangle_r = J_0$) are denoted by solid circles. Hence, LTP occurs in the range of ($\widetilde{D}_l^*$, $\widetilde{D}_h^*$); otherwise, LTD appears.
For comparison, the values of $\langle \langle J_{ij}^* \rangle \rangle_r$ for the additive case are also represented by crosses, and their lower and higher thresholds $\widetilde{D}_l$ $(\simeq 0.1179)$ and $\widetilde{D}_h$ $(\simeq 17.336)$ are denoted by stars. When passing $\widetilde{D}_l^*$, a transition to LTP occurs for the multiplicative case, and then $\langle \langle J_{ij}^* \rangle \rangle_r$ increases a little less rapidly, in
comparison with the rapid (step-like) transition for the additive case [see the left inset in Fig.~\ref{fig:MULTI1}(c)]. In the top region, a small ``plateau'' appears, then $\langle \langle J_{ij}^* \rangle \rangle_r$ decreases slowly [particularly, much slowly near the higher threshold $\widetilde{D}_h^*$ when compared with the additive case, as shown in the right inset in Fig.~\ref{fig:MULTI1}(c)], and a transition to LTD occurs as $\widetilde{D}_h^*$ is passed. Due to this relatively gradual transition, $\widetilde{D}_h^*$ becomes a little larger than $\widetilde{D}_h$. Hence, LTP for the multiplicative case occurs in a little wider range in comparison with the additive case. For most cases of LTP, the values of $\langle \langle J_{ij}^* \rangle \rangle_r$ are smaller than those for the additive case, due to soft bounds.

In addition to the population-averaged values $\langle \langle J_{ij}^* \rangle \rangle_r$, we are also concerned about the standard deviations $\sigma$ for distributions of $\{ J^*_{ij} \}$.
Figure~\ref{fig:MULTI1}(d) shows plots of $\langle \sigma \rangle_r$ versus $D$ for the multiplicative (represented by open circles) and the additive (denoted by crosses) cases;
the horizontal dotted line denotes the initial standard deviation $\sigma_0$ (=0.02), corresponding to case without STDP. As shown in histograms in Figs.~\ref{fig:MULTI1}(b1)-\ref{fig:MULTI1}(b6), standard deviations $\sigma$
for distributions of $\{ J^*_{ij} \}$ in the multiplicative case are much smaller than those for the additive case, due to the soft bounds for the multiplicative case. Moreover, the values of $\sigma$ for the multiplicative case are even smaller than $\sigma_0$ in the absence of STDP.

The effects of LTP/LTD on SBS may be well visualized in the raster plot of burst onset times. Figures \ref{fig:MULTI1}(e1)-\ref{fig:MULTI1}(e6) and Figures \ref{fig:MULTI1}(f1)-\ref{fig:MULTI1}(f6) show raster plots of burst onset times and their corresponding IPBR kernel estimates $R_b(t)$ for various values of $D$, respectively.
When compared with Figs.~\ref{fig:SBS1}(e1)-\ref{fig:SBS1}(e6) and Figs.~\ref{fig:SBS1}(f1)-\ref{fig:SBS1}(f6) in the absence of STDP, like the additive case, the degree of SBS for the case of LTP
($D=0.3,$ 5, 9, and 13) is increased so much due to increased $\langle \langle J_{ij}^* \rangle \rangle_r$, while in the case of LTD ($D=0.1175$ and 17.5) the population states become desynchronized due to
decreased $\langle \langle J_{ij}^* \rangle \rangle_r$.

For the case of LTP, we also make comparison with the additive case shown in Figs.~\ref{fig:STDP1}(d2)-\ref{fig:STDP1}(d5) and Figs.~\ref{fig:STDP1}(e2)-\ref{fig:STDP1}(e5).
As shown in Fig.~\ref{fig:MULTI1}(d), the standard deviations $\sigma$ of $\{ J_{ij}^* \}$ for the multiplicative case are much smaller than those for the additive case, although their values of the population-averaged
coupling strength $\langle \langle J_{ij}^* \rangle \rangle_r$ are also smaller (except for the case of $D=0.3$ where the values of $\langle \langle J_{ij}^* \rangle \rangle_r$ are nearly the same for both
the multiplicative and the additive cases). Effect of smaller $\sigma$ (increasing the degree of SBS) competes with effect of smaller $\langle \langle J_{ij}^* \rangle \rangle_r$ (decreasing the degree of SBS).
As a result, due to so much smaller standard deviations $\sigma$, the average maximum $\langle \langle R_b^{(max)} \rangle \rangle_r$ of the IPBR kernel estimate $R_b(t)$ becomes a little larger for the
multiplicative case, as shown in Fig.~\ref{fig:MULTI1}(g). It is not easy to directly compare the amplitudes of $R_b(t)$ for both the multiplicative and the additive cases in the scales of Figs.~\ref{fig:MULTI1}(f2)-\ref{fig:MULTI1}(f5) and Figs.~\ref{fig:STDP1}(e2)-\ref{fig:STDP1}(e5). Instead, in each realization, we obtain $\langle R_b^{(max)} \rangle$  via average over $3 \times 10^3$ global bursting cycles of $R_b(t)$ after the saturation time $t^*$ (= 500 sec for the multiplicative case and 2000 sec for the additive case), and $\langle \cdots \rangle_r$ represents an average over 20 realizations.
For the cases of LTP ($D=0.3$, 5, 9, and 13), the values of $\langle \langle R_b^{(max)} \rangle \rangle_r$ for the multiplicative case (denoted by open circles) are a little larger than those for the additive case (denoted by crosses). Consequently, as a whole, the degree of SBS for the multiplicative case seems to be a little higher than that for the additive
case, which will be discussed below in more details.

\begin{figure}
\includegraphics[width=0.7\columnwidth]{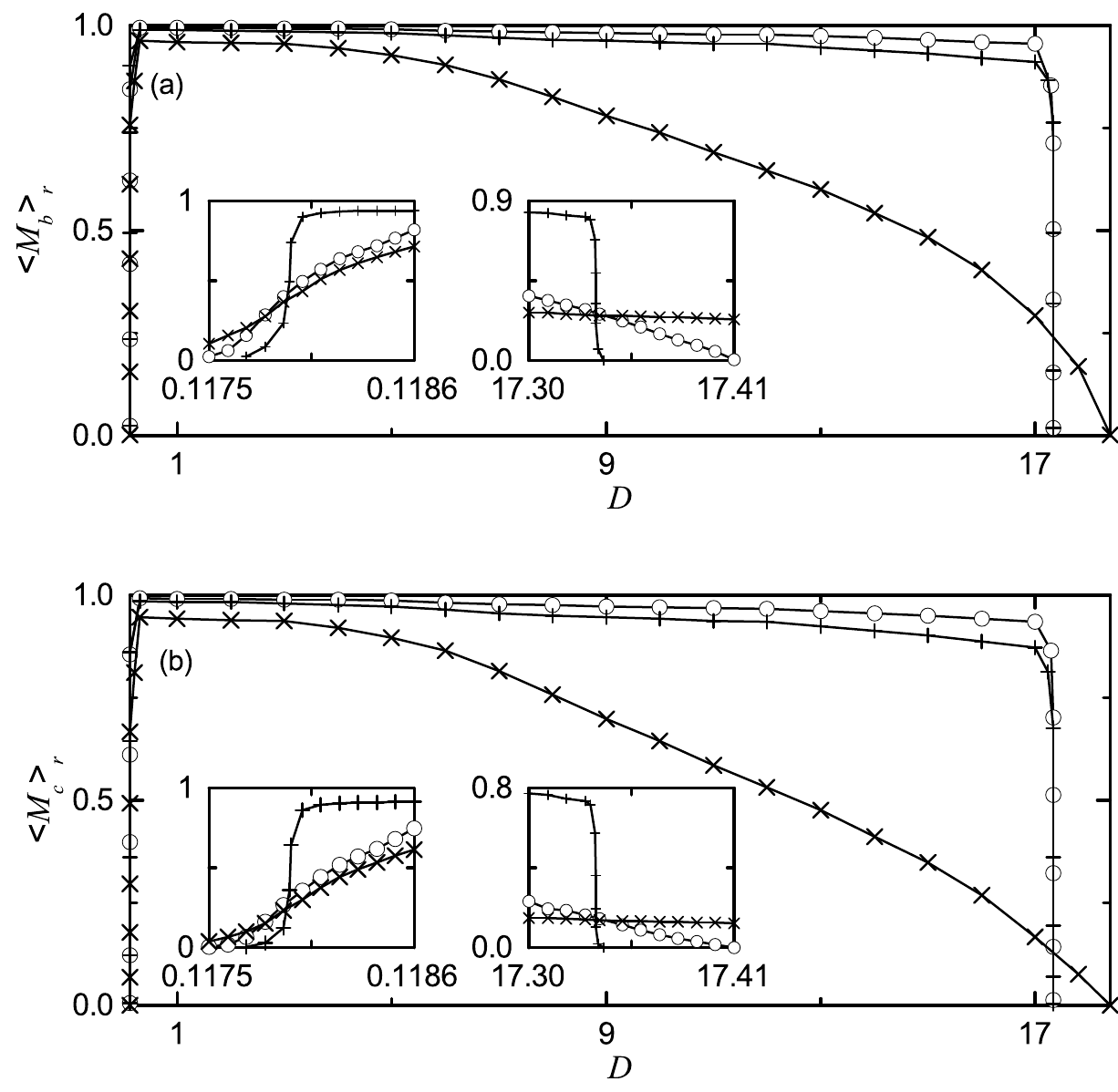}
\caption{Effects of multiplicative STDP on the statistical-mechanical bursting measure $M_b$ and the microscopic correlation measure $M_c$ for the case of symmetric attachment with $l^*=10$; $N=10^3$.
(a) Plot of $\langle M_b \rangle_r$ (represented by open circles) versus $D$; for comparison, $\langle M_b \rangle_r$ in the absence of STDP and for the case of additive STDP are also shown
in crosses and pluses, respectively. (b) Plot of $\langle M_c \rangle_r$ (denoted by open circles) versus $D$; for comparison, $\langle M_c \rangle_r$ in the absence of STDP and for the case of additive STDP are
also shown in crosses and pluses, respectively.
}
\label{fig:MULTI2}
\end{figure}

Finally, we investigate the effects of multiplicative STDP on the statistical-mechanical bursting measure $M_b$ of Eq.~(\ref{eq:BM}) and the microscopic correlation measure $M_c$ of Eq.~(\ref{eq:CM}).
Figure \ref{fig:MULTI2}(a) shows plots of $\langle M_b \rangle_r$ (denoted by open circles for the multiplicative case) versus $D$; for comparison, the values of $\langle M_b \rangle_r$ for the additive case and the case without STDP are also represented by pluses and crosses, respectively. Here, we get $\langle M_b \rangle_r$ by following $3 \times 10^3$ bursting stripes in the raster plot of burst onset times after the saturation time $t^*$ (=500 sec)
in each realization. Like the case of additive STDP, a Matthew effect in synaptic plasticity occurs via a positive feedback process, when compared with the static case without STDP. Good burst synchronization with higher $M_b$ gets better via LTP, while bad burst synchronization with lower $M_b$ gets worse via LTD. Consequently, a rapid transition to SBS occurs, in contrast to the relatively smooth transition in the absence of STDP.
However, due to soft bounds, changes near both ends are a little less rapid than those for the additive case, which are shown well in the insets of Fig.~\ref{fig:MULTI2}(a).
As a result of the effects of soft bounds, in most region of the top plateau in Fig.~\ref{fig:MULTI2}(a), the standard deviations $\sigma$ for the distribution of $\{ J_{ij}^* \}$ in the multiplicative case are much smaller than those for the additive case, although their population-averaged values $\langle \langle J_{ij}^* \rangle \rangle_r$ are also smaller (except for the case of small $D$ where $\langle \langle J_{ij}^* \rangle \rangle_r$ are nearly the same) [see Figs.~\ref{fig:MULTI1}(c) and \ref{fig:MULTI1}(d)]. Smaller standard deviation $\sigma$ (smaller $\langle \langle J_{ij}^* \rangle \rangle_r$) may increase (decrease) the degree of SBS. Since the effects of smaller standard deviations $\sigma$ are a little dominant, the values of $\langle M_b \rangle_r$ in most region of top plateau are a little larger than those for the additive case, in consistent with the results of $\langle \langle R_b^{(max)} \rangle \rangle_r$ in Fig.~\ref{fig:MULTI1}(g).
Figure \ref{fig:MULTI2}(b) shows plots of the microscopic correlation measure $\langle M_c \rangle_r$ for the multiplicative (``open circles'') and the additive (``pluses'') cases and
in the absence of STDP (``crosses''). The number of data used for the calculation of each temporal cross-correlation function $C_{ij}(\tau)$ [the values of $C_{ij}(0)$ at the zero time lag are used for calculation of $M_c$] is $2^{16}$ (=65536) after the saturation time $t^*$ (=500 sec) in each realization.
As in the case of $\langle M_b \rangle_r$, a Matthew effect also occurs in $\langle M_c \rangle_r$: good pair-correlation with higher $M_c$ gets better via LTP, while bad pair-correlation with lower $M_c$ gets worse via LTD. Hence, a rapid transition occurs, in contrast to the case without STDP. Like the case of $\langle M_b \rangle_r$, some quantitative differences arise, due to the effects of soft bounds.
Changes in $\langle M_c \rangle_r$ near both ends are a little less rapid than those for the additive case, which are shown well in the insets of Fig.~\ref{fig:MULTI2}(b).
In most region of the top plateau, the values of $\langle M_c \rangle_r$ for the case of multiplicative STDP are a little larger than those for the additive case, because the effects of smaller standard deviations (increasing the degree of pair-correlations) are a little dominant in comparison with the effects of smaller $\langle \langle J_{ij}^* \rangle \rangle_r$ (decreasing the degree of pair correlations).

\section{Summary}
\label{sec:SUM}
We considered an excitatory Barab\'{a}si-Albert SFN of subthreshold Izhikevich neurons which cannot fire spontaneously without noise. When the coupling strength passes a threshold, individual neurons
exhibit noise-induced burstings. We are concerned about SBS (i.e., population synchronization between noise-induced burstings) which may be an origin for synchronous brain rhythms in the noisy environment
which are correlated with brain function of encoding sensory stimuli. In our work, STDP for adaptive dynamics of synaptic strengths was taken into consideration, in contrast to previous works on the SBS
where synaptic strengths were static.

We first studied the effect of additive STDP (independent of states) by varying the noise intensity $D$ for the case of symmetric preferential attachment with the same in- and out-degrees ($l_{in}=l_{out}=l^*=10)$.
A Matthew effect in synaptic plasticity has been found due to a positive feedback process. Good burst synchronization (with higher bursting measure $M_b$) gets better via LTP of synaptic strengths, while bad burst
synchronization (with lower $M_b$) gets worse via LTD. As a result, a step-like rapid transition to SBS has been found to occur by changing $D$, in contrast to the relatively smooth transition in the absence of STDP.

In the presence of additive STDP, we have studied the effect of network architecture on SBS for a fixed $D~(=13)$ in the following two cases: (1) variations in (1) the symmetric attachment degree and (2) the
asymmetry parameter. For the first case of network architecture, as the symmetric attachment degree $l^*$ is increased from 10, the degree of SBS becomes better due to both better individual dynamics and better
efficiency of global communication between nodes (resulting from the increased number of total connections). On the other hand, with decreasing $l^*$ from 10, both individual dynamics and effectiveness of
communication between nodes become worse (resulting from the decreased number of total connections), and hence the degree of SBS becomes worse. In the second case of network architecture, with decreasing the asymmetry
parameter $\Delta l$ from 0, the degree of SBS becomes worse because both individual dynamics and efficiency of communication between nodes are worse. On the other hand, as $\Delta l$ is increased from 0, the degree of
SBS becomes better mainly because of better individual dynamics overcoming worse efficiency of communication.

We also investigated emergences of LTP and LTD of synaptic strengths intensively for the case of symmetric attachment with $l^*=10$ through our own microscopic methods based on both the distributions of time delays
$\{ \Delta t_{ij} \}$ between the pre- and the post-synaptic burst onset times and the pair-correlations between the pre- and the post-synaptic IIBRs. In the case of LTP, three (separate) peaks (a main central peak
and two left and right minor peaks) exist in the population-averaged histograms for the distributions of $\{ \Delta t_{ij} \}$, while a broad central peak appears through merging of the three peaks for the case of LTD.
Then, we could obtain population-averaged synaptic modifications $\langle \Delta J_{ij} \rangle$ from the population-averaged histograms, and they have been found to agree well with directly-calculated
$\langle \Delta J_{ij} \rangle$. Consequently, how microscopic distributions of $\{ \Delta t_{ij} \}$ contribute to $\langle \Delta J_{ij} \rangle$ may be clearly understood. Moreover, we studied the microscopic
correlation measure $M_c$, representing the in-phase degree between the pre- and the post-synaptic neurons, which are obtained from the pair correlations between the pre- and the post-synaptic IIBRs. As in the case
of bursting measure $M_b$, $M_c$ also exhibits a rapid transition due to a Matthew effect in the synaptic plasticity. Enhancement (suppression) of $M_c$ is directly related to decrease (increase) in the widths of bursting
stripes in the raster plot of burst onset times. Then, distributions of $\{ \Delta t_{ij} \}$ become narrow (wide), which may result in emergence of LTP (LTD). In this way, microscopic correlations between synaptic pairs
are directly related to appearance of LTP/LTD.

Furthermore, effects of multiplicative STDP (depending on states) on SBS have been investigated for the case of symmetric attachment with $l^*=10$ in comparison with the additive STDP case. Soft bounds for the
multiplicative case (i.e., a change in synaptic strengths scales linearly with the distance to the higher and the lower bounds) are in contrast to hard bounds for the additive case. Some quantitative differences
between the results for the additive and the multiplicative STDP arise because of the soft bounds. As in the case of additive STDP, a Matthew effect has been found to occur in the bursting measure $M_b$. However, due to
the soft bounds, a relatively less rapid transition occurs near both ends, in comparison to the rapid transition for the additive cases. Moreover, due to the soft bounds, the standard deviations $\sigma$ for the
distributions of saturated limit synaptic strengths $\{ J^{*}_{ij} \}$ are much smaller than those for the additive case. As a result of the smaller standard deviations $\sigma$, the degrees of SBS (given by $M_b$)
in most plateau-like top region (corresponding to most cases of LTP) become a little larger than those in the additive case. A Matthew effect has also been found to occur in the microscopic correlation measure $M_c$.
Good pair-correlation (with higher $M_c$) gets better via LTP, while bad pair-correlation (with lower $M_c$) gets worse via LTD. However, like the case of $M_b$, some quantitative differences in $M_c$ (for the additive
and the multiplicative STDP) also occur near both ends and in most plateau-like top region, due to the soft bounds.

Finally, we briefly discuss relevant ones associated with our work (e.g., biological implication, other neuronal models, and other measures).
Our simulation work on the effect of STDP on SBS is closely related to neuroscience because the STDP controls the efficacy of the brain function of encoding sensory stimuli in the noisy environment mediated by
the bursting neurons (e.g. in cortex, thalamus, hippocampus, or cerebellum) in the complex neuronal network. As explained in Sec.~\ref{sec:SFN}, the Izhikevich neuron model in our work is biologically plausible,
as in the Hodgkin-Huxley-type conductance-based models, and hence we expect that our results would be valid in other biological models such as the Hindmarsh-Rose \cite{HR1,HR2,HR3} and the Hodgkin-Huxley \cite{HH1} models.
For characterization of burst synchronization, we employed a statistical-mechanical bursting measures $M_b$ which, in a statistical-mechanical way, measures the occupation (representing the density of bursting stripes) and
the pacing (denoting the smearing of bursting stripes) degrees of burst synchronization, visualized well in the raster plot. Our statistical-mechanical bursting measure $M_b$ is in contrast to the conventional microscopic
burst phase order parameter $r$ \cite{measure1,measure2} because $r$ quantifies the coherence degree between microscopic individual burst phases without any explicit relation to the macroscopic occupation and pacing
patterns of burst onset times visualized well in the raster plot. In a statistical-mechanical sense, our bursting measure $M_b$ supplements the conventional microscopic measure $r$. Hence, instead of $M_b$, one may use
the conventional microscopic burst phase order parameter $r$, and the same results are expected for characterization of burst synchronization.
We also make brief description on future works.
In the present work, we investigated the effect of an excitatory STDP on SBS in an excitatory population. SBS was also found to occur in an inhibitory population \cite{Kim2}. Hence, it would be interesting to study the effect
of inhibitory STDP on SBS. However, inhibitory STDP was less studied due to experimental obstacles and diversity of inhibitory interneurons \cite{iSTDP1}. The inhibitory population was also found to exhibit diverse non-Hebbian
inhibitory STDP \cite{iSTDP2,iSTDP1,Tass1}, in contrast to the case of excitatory Hebbain STDP. Since the work on the inhibitory STDP is beyond the present work, it is left as a future work.
The additive STDP (update) rule is independent of states, while the multiplicative STDP rule depends on states. Particularly, as a multiplicative STDP rule, we consider a linearly-dependent case [see Eq.~(\ref{eq:MSTDP})].
There exists another multiplicative STDP rule with nonlinear power-law dependence \cite{PowerSTDP}. When the exponent of the power law is 1, it corresponds to our multiplicative case with linear dependence, while it approaches
the additive case as the exponent goes to the zero. Hence, in future, it seems to be interesting to investigate the nonlinear multiplicative case by changing its exponent from 1 to 0, and compare the results with those for both
the additive and the linear multiplicative cases.
Brain has a modular clustered structure which may be modelled as a clustered SFN. Hubs in each cluster are strongly interconnected, and they form a rich club for effective global communication via integration of neural
information in diverse brain modules \cite{rich1,rich2}. Hence, it would be interesting to study the rich-club effect on STDP in a clustered SFN. However, it is beyond the present work, and hence it is left as a future work.

\section*{Acknowledgments}
This research was supported by the Basic Science Research Program through the National Research Foundation of Korea (NRF) funded by the Ministry of Education
(Grant No. 20162007688).

\end{document}